\documentclass[final,1p,times,authoryear]{elsarticle}
\usepackage{amssymb}
\usepackage{amsmath}
\usepackage{amsthm}
\usepackage{url}
\usepackage{float}
\usepackage{algorithm}
\usepackage{algpseudocodex}
\usepackage{csquotes}
\newtheorem{theorem}{Theorem}
\usepackage{graphicx}
\usepackage{caption}
\usepackage{subcaption}
\usepackage{multirow}
\usepackage{array}
\usepackage{doi}

\makeatletter
\renewcommand{\thealgorithm}{\arabic{algorithm}}
\newif\if@appendix
\let\oldappendix\appendix
\renewcommand{\appendix}{\oldappendix\@appendixtrue\renewcommand{\thealgorithm}{\Alph{section}.\arabic{algorithm}}}
\makeatother

\journal{Computational Statistics \& Data Analysis}

\begin{document}

\begin{frontmatter}

\title{rfBLT: Random Feature Bayesian Lasso Takens Model\\ for time series forecasting} 

\author{Thu Nguyen, Lam Si Tung Ho} 

\affiliation{organization={Dalhousie University},
            addressline={6297 Castine Way - PO Box 15000}, 
            city={Halifax},
            postcode={B3H 4R2}, 
            state={Nova Scotia},
            country={Canada}}

\begin{abstract}
Time series prediction is challenging due to our limited understanding of the underlying dynamics. Conventional models such as ARIMA and Holt's linear trend model 
experience difficulty in identifying nonlinear patterns in time series. In contrast, machine learning models excel at learning complex patterns and handling high-dimensional 
data; however, they are unable to quantify the uncertainty associated with their predictions, as statistical models do. To overcome these drawbacks, we propose Random Feature 
Bayesian Lasso Takens (rfBLT) for forecasting time series data. This non-parametric model captures the underlying system via the Takens' theorem and measures the degree of 
uncertainty with credible intervals. This is achieved by projecting delay embeddings into a higher-dimensional space via random features and applying regularization within 
the Bayesian framework to identify relevant terms. Our results demonstrate that the rfBLT method is comparable to traditional statistical models on simulated data, while 
significantly outperforming both conventional and machine learning models when evaluated on real-world data. 
The proposed algorithm is implemented in an R package, named \texttt{rfBLT}.
\end{abstract}


\end{frontmatter}

\section{Introduction}
Forecasting time series is a crucial task in many areas; an accurate prediction model could provide beneficial insights for proactive planning, helping to avoid undesirable events. Specifically, they may provide information about the distribution of resources; for instance, during the COVID-19 pandemic, forecasting new cases could have enabled governments to more effectively distribute resources such as vaccines and medical personnel across various regions. Similarly, the successful projection of stock prices could be helpful for economic planning by offering insights about market trends and potential opportunities.

However, predicting these observable variables is difficult due to our incomplete understanding of their underlying systems. 
Furthermore, the scarcity of data regarding these variables is also a significant challenge in modeling these time series. 
Traditional models, like Autoregressive Integrated Moving Average (ARIMA) and Exponential Smoothing, do not require assumptions about the underlying causal relationships of the system, beyond the statistical assumptions about the observed time series itself. 
Nevertheless, in practice, causal connections exist between different time series. For instance, the increase in cases during the COVID-19 pandemic was followed by a rise in deaths. 
Therefore, it is necessary to build models capable of forecasting observed time series variables despite limited system knowledge and shortage of data.

Time delay embeddings are well-known for reconstructing dynamical systems \citep{mcgoff2015statistical}. \citet{kaiser2018sparse} proposed the Sparse Identification of 
Nonlinear Dynamics (SINDy) algorithm, which utilizes regularization regression with respect to nonlinear functions, such as polynomials and trigonometric functions, on 
time delay embeddings to identify nonlinear dynamics based on measurement data. They assume that the dynamics are governed by just a few key terms, which are applied in many systems. 
In epidemiology, \citet{saha2023spade4} proposed the Sparsity and Delay Embedding Based Forecasting of Epidemics (SPADE4) model. 
This approach leverages Takens' delay embedding theorem to capture underlying system patterns, and employs random feature models (RFMs) to map data into a higher-dimensional space using randomized nonlinear functions, as opposed to fixed nonlinear mappings like trigonometric or polynomial functions.
Their results demonstrate the superiority of the proposed method compared to benchmark models on both simulated and real data. 
In addition, they suggest that the random feature model with penalized regression helps to effectively learn complex patterns while avoiding overfitting caused by limited data. 
This has made sparsity RFMs gain significant research interest in recent years \citep{saha2023harfe, richardson2024srmd}.

Conceptually, RFMs operate as a class of two-layer neural networks with fixed weights in the first layer \citep{liu2021random}.
This approach offers a more compelling alternative to classical neural networks because, while neural networks can produce highly accurate results with well-tuned parameters, they require a high computational cost for training models with a large number of predictors. 
Additionally, conventional neural networks do not naturally provide estimates of predictive uncertainty \citep{gawlikowski2023survey}.
First introduced by \citet{rahimi2007random}, RFMs provide an alternative approach to approximate shift-invariant kernels using randomized feature maps, while reducing computational cost. 
Other works of \citet{rudi2017generalization, mei2022generalization, mei2022rfreg, hashemi2023generalization} explored the theoretical guarantees of generalization error bounds of sparse random features. 

In high-dimensional settings, where the number of dimensions exceeds the number of data points, a typical scenario in RFMs, the implementation of feature selection techniques is critical to avoid overfitting. 
Penalized regression is commonly known for its robustness to model noise and prevents overfitting by shrinking the irrelevant coefficients toward 0 \citep{hoerl1970ridge, tibshirani1996regression, zou2003regression}. 
The renowned package \texttt{glmnet} implements the ridge, lasso, and elastic net regularization regressions \citep{friedman2010regularization}. 
While these are frequentist approaches, the concept was also applied in Bayesians, in which the shrinkage prior distributions are determined for coefficients corresponding to different kinds of regularization. 
The priors of coefficients are widely studied by \citet{hsiang1975bayesian, park2008bayesian, li2010bayesian, van2019shrinkage}. 
A state-of-the-art software toolbox to implement penalized regression within the Bayesian framework for linear models is developed by \citet{makalic2016high} in both {R} and {MATLAB}, named \texttt{bayesreg}.  

In this work, we propose a statistical method named Random Feature Bayesian Lasso Takens (rfBLT) to predict future values of time series and provide credible intervals to 
quantify the uncertainty of predictions. rfBLT is a cooperation of the Random Feature Model and the Time delay embedding theorem under 
the Bayesian framework. Specifically, we map the embedding time series to another dimensional space using random features, then apply the penalized regression to model the 
approximation of the smoothness rate of change using the Bayesian approach. 
The proposed model provides uncertainty quantification despite a limited amount of data via the sparse Bayesian regression and random feature maps, an advantage over standard neural networks.
It also accounts for prior information, which the frequentist approach in the work of \citet{saha2023spade4} does not.

Our goal is to predict and capture the trend of the trajectories by evaluating the 7-day-ahead recursive forecasts. 
We analyze and compare our model performance to a variety of other models, including Random Feature Bayesian Lasso without utilizing Takens' theorem (rfBL), traditional statistical time series models 
(ARIMA and Holt's linear trend model), and machine learning models (Long Short-Term Memory (LSTM) and Random Forest). 
These models are applied to both the simulated proportional cases (produced by the compartmental epidemiological model (S$\mu$EIR), proposed by \citet{zou2020epidemic}, with 10\% noise level)
and real-world time series, including the S\&P 500 daily closing price, and the COVID-19 daily new cases and 
deaths in Canada. In the context of simulations, our proposed model yields results comparable to classical statistical models, while it outperforms all other methods in 
real-world datasets in terms of relative errors, mean directional accuracy, and coverage. We implement the proposed model in the R package \texttt{rfBLT}, which can be found at
\href{https://github.com/thuthiminhnguyen/rfBLT}{github.com/thuthiminhnguyen/rfBLT}.  

The paper is organized as follows. Section~\ref{sec_method} describes the proposed model and its components. Section~\ref{sec_baseline} outlines benchmark models. 
Section~\ref{sec_experiment_setup} presents simulated and real datasets, provides the configuration setup of each algorithm for each dataset, and introduces measurement 
metrics. Section~\ref{sec_experiment_res} compares and contrasts outcomes from various models. Section~\ref{sec_conclusion_discuss} concludes and discusses the key findings. 

\section{Methodology}\label{sec_method}
\textbf{Notation.} This paper uses bold letters for column vectors (e.g., $\boldsymbol{\beta}$) and bold capital letters for matrices (e.g., $\mathbf{W}$). 
Time differentiation is represented with the prime symbol (e.g., $y'$), the smoothed value is denoted with the bar symbol (e.g., $\bar{y}$), 
and the predicted value is marked with the hat (e.g., $\hat{y}$). 
\subsection{Random Feature Model}
The kernel trick is a powerful technique applied to handle non-linear relationships between data points with efficient computations in kernel methods. 
It not only simplifies computations but also enhances the comprehensiveness of models \citep{10.1214/009053607000000677, ghojogh2019unsupervised}. 
Conventionally, when there are $n$ data points, the algorithms need to calculate the kernel matrix of size $n\times n$ with each entry $K_{ij}=\kappa(\mathbf{x}_i, \mathbf{x}_j),$ where $\mathbf{x}_i\in\mathbb{R}^p$ and $i, j=1, \dots, n.$  
This approach leads to storage and computation requirements of $\mathcal{O}(n^2)$ and $\mathcal{O}(n^3)$ for training, respectively, which is computationally costly for large datasets. 
To address this issue, \citet{rahimi2007random} introduced Random Feature Models (RFMs) to approximate kernel matrices with reduced computational cost. 
RFMs reduce the cost of storage to $\mathcal{O}(nD)$ and training time to $\mathcal{O}(npD)$, where $D$ is the number of random features.

Consider the following response data $\mathbf{y}=[y_1, \dots, y_n]^T\in\mathbb{R}^n$ with the input matrix of  $\mathbf{X}\in \mathbb{R}^{n\times p}$, the random feature algorithm constructs a randomized feature map from $\mathbb{R}^p$ to $\mathbb{R}^D$. 
The activation function $\sigma(\cdot)$ is parameterized by the weighting matrix $\mathbf{W}\in \mathbb{R}^{p\times D}$ and the bias vector $\mathbf{b}\in \mathbb{R}^{D}$, which are randomly drawn from the specified probability distributions $p(\mathbf{W})$ and $p(\textbf{b})$, respectively \citep{rahimi2008weighted}. 
The linear combination function of $\mathbf{y}$ using random feature model can be expressed as 
\begin{equation}\label{rfm_eq1}
    y_i=\beta_0+\sigma(\mathbf{x}_i^T\mathbf{W}+\mathbf{b}^T)\boldsymbol{\beta}+\epsilon_i,
\end{equation}
where $\mathbf{x}_i\in\mathbb{R}^p$, $\beta_0$ is the estimated intercept, $\sigma(\cdot)$ is the activation function, $\boldsymbol{\beta}\in \mathbb{R}^D$ is the learned weights such that the loss function is minimized, and $\epsilon_i$ is the error term. Alternatively, defining $\mathbf{z}_i=\sigma(\mathbf{x}_i^T\mathbf{W}+\mathbf{b}^T)\in\mathbb{R}^D$ as the transformed random feature mapping, the model can be expressed as a linear regression in the feature space as follows:
\begin{equation}\label{rfm_eq2}
    g(\mathbf{z}_i)=\beta_0+\mathbf{z}_i^T\boldsymbol{\beta}+\epsilon_i,
\end{equation}
or in the matrix form of
\begin{equation*}
    g(\mathbf{Z})=\beta_0+\mathbf{Z}\boldsymbol{\beta}+\boldsymbol{\epsilon}.
\end{equation*}

The random weighting matrix $\mathbf{W}$ and bias vector $\mathbf{b}$ can be sampled from various distributions, such as uniform, normal, Cauchy, exponential, Bernoulli, and lognormal, each with specified arguments, which are provided in the \texttt{rfBLT} package. The common activation functions are also supported, including
\begin{enumerate}
    \item Fourier activation function: $\sqrt{\frac{2}{D}}\cos(\mathbf{x}_i^T\mathbf{W}+\mathbf{b}^T)$. 
    \item ReLU activation function: $\max(0, \mathbf{x}_i^T\mathbf{W}+\mathbf{b}^T)$.
    \item Sigmoid activation function: $\frac{1}{1+e^{-(\mathbf{x}_i^T\mathbf{W}+\mathbf{b}^T)}}$.
    \item Tanh activation function: $\frac{e^{(\mathbf{x}_i^T\mathbf{W}+\mathbf{b}^T)} - e^{-(\mathbf{x}_i^T\mathbf{W}+\mathbf{b}^T)}}{e^{(\mathbf{x}_i^T\mathbf{W}+\mathbf{b}^T)} + e^{-(\mathbf{x}_i^T\mathbf{W}+\mathbf{b}^T)}}$.
    \item Sine activation function: $\sin(\mathbf{x}_i^T\mathbf{W}+\mathbf{b}^T)$.
    \item Cosine activation function: $\cos(\mathbf{x}_i^T\mathbf{W}+\mathbf{b}^T)$.
\end{enumerate}

\paragraph{Number of features} A natural question arises: how many features are enough to produce an efficient predictive model?

For large datasets with $n$ observations and $p$ dimensions, to ensure computational efficiency, it is common to 
select a number $D$ of random features that falls between $p$ and $n$ \citep{liu2021random}. 
Initially, \citet{rahimi2008weighted}  found that $n$ features are enough to ensure no loss of learning accuracy 
for models utilizing Random Fourier Features (RFF) with Lipschitz continuous functions. 
However, the number of feature choices varied among studies, as each aimed to achieve different error bounds. 
For instance, research on the generalization properties of ridge regression with random features under a fixed Reproducing Kernel Hilbert Space (RKHS) shows that 
models can achieve the error of order $1/\sqrt{n}$ with $\sqrt{n}\log(n)$ features, which is the same prediction accuracy as the exact kernel ridge regression estimator \citep{rudi2017generalization}. 
On the other hand, \citet{mei2022generalization}, using a different treatment for the target function, 
demonstrated that the error of random features ridge regression models is comparable to that of kernel ridge regression, provided the number of features of $D\geq n^{1+\gamma}$ for some $\gamma>0$. 
These findings suggest that there is no precise number of features; the optimal choice depends on the problem's characteristics or the need to balance model accuracy with complexity. Hence, the \texttt{rfBLT} package provides flexible options for setting the number of features $(D)$, such as $\sqrt{n}$, a multiplier of $n$, and a fixed number of random features. 

\subsection{Bayesian Regularization Regression}
To prevent overfitting and facilitate variable selection in predictive modeling, the frequentist approach for the regularization regression incorporates penalty terms. 
Bayesian regularization regression, however, estimates parameters by leveraging their prior distributions and calculating the posterior distributions. Both approaches 
shrink coefficients with minor effects towards zero and preserve those with substantial effects \citep{van2019shrinkage}. The \texttt{rfBLT} package includes implementations 
of Bayesian lasso and Bayesian ridge for RFMs. For these, the posterior distributions of all parameters are computed using the \texttt{bayesreg::bayesreg} function, 
which supports efficient Gibbs sampling for high-dimensional data \citep{makalic2016high}. The \texttt{rfBLT} package also implements sparsity RFMs within the frequentist 
framework. It supports ridge, lasso, and elastic net regularization, leveraging coefficients estimated using the \texttt{glmnet} package \citep{friedman2010regularization}.

In RFMs, the predictor vector $\mathbf{z}_i = \sigma(\mathbf{x}_i^T\mathbf{W} + \mathbf{b}^T) \in \mathbb{R}^{D}$ represents features transformed via an activation function $\sigma(\cdot)$. Assuming Gaussian errors, the model from Equation \ref{rfm_eq2} is represented as:
\begin{equation*}
    y_i|\mathbf{z}_i, \beta_0, \boldsymbol{\beta}, \sigma^2_\epsilon\sim\mathcal{N}(\beta_0+\mathbf{z}^T_i\boldsymbol{\beta}, \sigma_\epsilon^2),
\end{equation*}
or in the matrix form of 
\begin{equation*}
    \mathbf{y}|\mathbf{Z}, \beta_0, \boldsymbol{\beta}, \sigma_\epsilon^2\sim \mathcal{N}(\beta_0+\mathbf{Z}\boldsymbol{\beta}, \sigma_\epsilon^2\mathbf{I}),
\end{equation*}
where $(\beta_0, \boldsymbol{\beta}, \sigma_\epsilon^2)$ are the parameters of interest, and $\mathbf{I}$ is an $n\times n$ identity matrix.
The corresponding log likelihood can be derived as follows:
\begin{equation*}
    \log p(\mathbf{y}|\boldsymbol{\beta}, \beta_0, \sigma_\epsilon^2) = \sum_{i=1}^n\left[\log\left(\frac{1}{\sqrt{2\pi}\sigma_\epsilon}\right)-\frac{(y_i-\beta_0-\mathbf{z}_i^T\boldsymbol{\beta})^2}{2\sigma_\epsilon^2}\right]
    = n\log\left(\frac{1}{\sqrt{2\pi}\sigma_\epsilon}\right)-\frac{1}{2\sigma_\epsilon^2}\|\mathbf{y}-\beta_0\mathbf{1}_n-\mathbf{Z}\boldsymbol{\beta}\|^2_2,
\end{equation*}
where $\mathbf{1}_n\in\mathbb{R}^n$ is a column vector and $\mathbf{z}_i=[z_{i1}, z_{i2}, \dots, z_{iD}]^T$. 

In the frequentist framework, the penalty term is added to the residual sum of squares to shrink the coefficients with less influence towards 0 and keep the coefficients with significant influence. In particular, ridge regression reduces coefficients to but not equal zero, whereas lasso regression can shrink less significant coefficients to precisely zero, encouraging sparsity. The coefficients $\beta_0$ and $\boldsymbol{\beta}$ are estimated by minimizing the following function
\begin{equation*}
    \|\mathbf{y}-\beta_0\mathbf{1}_n-\mathbf{Z}\boldsymbol{\beta}\|_2^2+\lambda\|\boldsymbol{\beta}\|_q,
\end{equation*}
where $\lambda$ is the penalty tuning parameter, and $q$ determines the penalty type: $q=1$ for lasso regression, and $q=2$ for ridge regression.

In the Bayesian framework, shrinkage is determined using the prior distributions of the regression coefficients with the specified hyperparameters. The hierarchical regression model with Gaussian errors is defined as follows:
\begin{align*}
y_i&\sim\mathcal{N}(\beta_0+\mathbf{z}^T_i\boldsymbol{\beta}, \sigma_\epsilon^2), \\
\beta_0 & \sim \pi(\beta_0) \propto 1, \\
\beta_j | \sigma_\epsilon^2, \tau^2, \lambda_j^2 & \sim \mathcal{N}(0, \lambda_j^2 \tau^2 \sigma_\epsilon^2), \\
\sigma_\epsilon^2 & \sim \pi(\sigma_\epsilon^2) \propto \sigma_\epsilon^{-2}, \\
\tau^2 &\sim \pi(\tau^2),\\
\lambda_j^2 & \sim \pi(\lambda_j^2),
\end{align*}
where $i=1, \dots, n$, $j=1, \dots, D$, the improper uniform prior for $\beta_0$ and scale-invariant prior for $\sigma_\epsilon^2$ ensure non-informative inference. The different prior distributions can be defined via the hyperparameters $\tau^2$ (global shrinkage hyperparameter) and $\lambda_j$ (local shrinkage hyperparameter) that encourage sparsity in the regression coefficients. The \texttt{bayesreg} package employs efficient Gibbs sampling, leveraging algorithms of \citet{rue2001fast} for $D/n < 2$, and \citet{bhattacharya2016fast} otherwise, ensuring scalability for RFMs. 

In the \texttt{bayesreg} package, the prior of $\tau$ is defined as
\begin{align*}
    \tau^2|\xi & \sim \text{Inv-Gamma}(1/2,1/\xi), \\
    \xi &\sim \text{Inv-Gamma}(1/2, 1),
\end{align*}
where $\xi>0$ is a mixing parameter, and Inv-Gamma$(a, b)$ is the Inverse Gamma distribution with the shape and scale parameters of $a$ and $b$, respectively. This leads to the posterior of $(\beta_0, \boldsymbol{\beta}, \sigma_\epsilon^2, \tau^2, \lambda^2)$ as follows:
\begin{align*}
    p(\beta_0, \boldsymbol{\beta}, \sigma_\epsilon^2, \tau^2, \xi,\lambda^2|\mathbf{Z}, {\mathbf{y}})&\propto p({\mathbf{y}}|\beta_0, \boldsymbol{\beta}, \sigma_\epsilon^2, \tau^2, \xi, \lambda^2, \mathbf{Z})p(\beta_0|\boldsymbol{\beta}, \sigma_\epsilon^2, \tau^2, \xi, \lambda^2)p(\boldsymbol{\beta}| \sigma_\epsilon^2, \tau^2, \xi, \lambda^2)\\
    & \hspace{1cm}p(\sigma_\epsilon^2|\tau^2, \xi,\lambda^2)p(\tau^2|\xi, \lambda^2)p(\xi|\lambda^2)p(\lambda^2)\\
    &= p({\mathbf{y}}|\beta_0, \boldsymbol{\beta}, \sigma_\epsilon^2, \tau^2, \xi, \lambda^2, \mathbf{Z})p(\beta_0)p(\boldsymbol{\beta}|\sigma_\epsilon^2, \tau^2, \lambda^2)p(\sigma_\epsilon^2)p(\tau^2|\xi)p(\xi)p(\lambda^2).
\end{align*}


\noindent For Bayesian ridge regression, the prior distribution of coefficients is a zero-mean Gaussian distribution, and $\lambda_j$ is set to 1 for all $j=1, \dots, D$. These give
\begin{equation*}
    \beta_j|\sigma_\epsilon^2, \tau^2 \sim \mathcal{N}\left(0,\tau^2\sigma_\epsilon^2\right).
\end{equation*}


\noindent For Bayesian lasso regression, the \texttt{bayesreg} package defines the hierarchy differently from the original work of \citet{park2008bayesian}. 
Particularly, a Gaussian variance mixture distribution with an exponential distribution as the mixing density is used to replace the Laplace distribution. 
The hierarchical model is updated with the following components:  
\begin{equation*}
    \beta_j|\sigma_\epsilon^2, \tau^2, \lambda_j^2 \sim \mathcal{N}(0, \lambda_j^2\tau^2\sigma_\epsilon^2), \hspace{0.3cm}\lambda_j^2\sim\text{Exp}(1).
\end{equation*}

\noindent Further details on the algorithm’s implementation for ridge and lasso can be found in \ref{app:bayesian_ridge_gibbs} and \ref{app:bayesian_lasso_gibbs}, respectively.

\subsection{Time Delay Embedding}
The Takens' theorem, or delay embedding theorem, was developed by Floris Takens in 1981. Under the assumption that the system is deterministic, time-delay embedding provides a way to reconstruct the state of the system from observed time series \citep{takens2006detecting}. 

\begin{theorem}[Takens' Theorem]\label{taken_theo}
Let M be a compact manifold of dimension d. For pairs $(\varphi, \text{y})$, $\varphi:$ M $\to$ M a smooth diffeomorphism and $\text{y}:$ M$\to\mathbb{R}$ a smooth function, it is a generic property that the map $\Phi_{(\varphi, \text{y})}:$ M$\to\mathbb{R}^{2d+1}$, defined by
\begin{equation*}
    \Phi_{(\varphi, \text{y})}(\text{x})=(\text{y(x)}, \text{y}(\varphi(\text{x})), \dots, \text{y}(\varphi^{2d}(\text{x})))
\end{equation*}
is an embedding.
\end{theorem}
\noindent Particularly, the term ``generic" means open and dense, and ``smooth" means at least twice continuously differentiable. 

\vspace{0.2cm}
An embedding is defined as a mapping that preserves the differential structure of the original space \citep{sauer1991embedology}. Suppose that the derivative map or the Jacobian matrix of $\varphi$ at $x$ ($\mathbf{J_\varphi}$) is one-to-one at every point $x$ of $M$, then the smooth map $\varphi$ on $M$ is an immersion. This implies that $\mathbf{J_\varphi}$ has full rank on the tangent space, and smooth observables (e.g., time derivatives) can be reconstructed since the tangent space structure is preserved. 
In addition, the underlying dynamics can be approximated by observing how the delay coordinates of a given point relate to the system's state after $T$ time units \citep{eckmann1985ergodic}. Consequently, the modeled time delay vectors can be used to reconstruct the dynamics.

The forward Euler method, commonly referred to as the Euler method, is a first-order numerical approach for solving ordinary differential equations (ODEs) with a specified initial condition. Consider a time series with length $n$, denoted by $\{(y_{t_k})\}_{k=1}^n$, the forward difference approximation implemented in \texttt{rfBLT} is defined as:
\begin{equation}\label{fwd_euler_bayes}
    y'_{t_k} \approx \frac{y_{t_{k+1}}-y_{t_k}}{t_{k+1}-t_k},
\end{equation}
where $k=1, \dots, n-1$. From this, $y_{t_{k+1}}$ can be expressed as:
\begin{equation}\label{fwd_diff_pred}
    y_{t_{k+1}}\approx y_{t_k}+y'_{t_k}(t_{k+1}-t_k).
\end{equation}

Since finite difference methods can yield unreliable results when applied to noisy data, a denoising method can be employed to capture the underlying trend \citep{saha2023spade4}. In this work, the left-moving average for derivative smoothing is applied, which is defined as follows: 
\begin{equation}\label{avg_fwd_euler_bayes}
    \bar{y}'_{t_k}=
    \begin{cases}
        \frac{1}{k}\sum_{j=1}^ky'_{t_j}, & k=1, \dots, s-1,\\
        \frac{1}{s}\sum_{j=k-s+1}^ky'_{t_j}, & k=s, \dots, n-1,
    \end{cases}
\end{equation}
where $s$ represents the level of smoothing such that $1\leq s\leq n$. There are other methods for obtaining the smoothness which are implemented in the \texttt{rfBLT} package, such as polynomial smoothing, low-pass filtering, locally estimated scatterplot smoothing (LOESS), and spline smoothing.

Additionally, we assume that the error between the smoothed derivative and the derivative obtained from Equation~\ref{fwd_euler_bayes} is normally distributed, which can be expressed as:
\begin{equation*}
    \delta_k=y'_{t_k}-\bar{y}'_{t_k}\sim\mathcal{N}(0, \sigma_{\delta}^2),
\end{equation*}
where $k=1, \dots, n-1$. By using Takens' Theorem~\ref{taken_theo}, the time derivatives can now be modeled by the time delay vectors themselves, which are determined by
\begin{equation*}
    \bar{y}'_{t_k}=f(y_{t_k}, y_{t_{k-1}}, \dots, y_{t_{k-m+1}}),
\end{equation*}
where $k\geq m$.
\subsection{Random Features Bayes Lasso Takens (rfBLT)}\label{sec_rfBLT}

The rfBLT method aims to capture the nonlinear relationship within the embedding time frame. 
Specifically, it predicts the smooth time derivatives $(\bar{\mathbf{y}}')$, which are then used to forecast future time series values using Equation~\ref{fwd_diff_pred} while correcting for errors.

Given a time series $\{(y_{t_k})\}_{k=1}^n$, the time derivatives $({y}'_{t_k})$ are then derived using Equation~\ref{fwd_euler_bayes}. Subsequently, the smoothness rate of change $(\bar{y}'_{t_k})$ is obtained using Equation~\ref{avg_fwd_euler_bayes}. The corresponding input-output pairs are organized as follows:
\begin{equation*}
    \mathbf{X}\gets
    \begin{bmatrix}
        y_{t_1} & y_{t_2} & \dots & y_{t_{m-1}} & y_{t_m}\\
        y_{t_2} & y_{t_3} & \dots & y_{t_m} & y_{t_{m+1}}\\
        \vdots & \vdots & & \vdots & \vdots\\
        y_{t_{n-m}} & y_{t_{n-m+1}} & \dots & y_{t_{n-2}} & y_{t_{n-1}}
    \end{bmatrix}=\begin{bmatrix}
        \mathbf{x}_m\\
        \mathbf{x}_{m+1}\\
        \vdots\\
        \mathbf{x}_{n-1}
    \end{bmatrix}    
    , \hspace{0.5cm}
    \bar{\mathbf{y}}'\gets\begin{bmatrix}
        \bar{y}'_{t_m}\\
        \bar{y}'_{t_{m+1}}\\
        \vdots\\
        \bar{y}'_{t_{n-1}}
    \end{bmatrix},
\end{equation*}
where $m$ is the embedding window size.

The rfBLT algorithm is summarized in Algorithm~\ref{alg:RaFBLT_EC} with the following parameters of interest: the intercept ($\beta_0$), coefficients ($\boldsymbol{\beta}$), and the error variance ($\sigma_\epsilon^2$). 
In addition, the recursive forecasting is applied to predict future values $h$ days ahead. 

\begin{algorithm}[H]
\caption{Random Feature Bayesian Lasso Takens (rfBLT) for Time-Series Data}
\label{alg:RaFBLT_EC}
\begin{algorithmic}[1]
\Require{Input time series $\{(y_{t_k})\}_{k=1}^n$; smoothing function $f_{\text{smooth}}(\cdot)$; embedding dimension $m$; number of features $D$; distribution for sampling weighting matrix $p(\mathbf{W})$; distribution for sampling bias vector $p(\mathbf{b})$; activation function $\sigma(\cdot)$; number of samples $S$; number of burn in samples ${B}$; prediction horizon $h$; credible interval level $(1-\alpha)100\%$.}
\Ensure{Predicted values $\{(\hat{y}^{\text{mean}}_{t_k})\}_{k=n+1}^{n+h}$ and $(1-\alpha)100\%$ credible intervals $\{\left[\hat{y}^{\text{lower}}_{t_k}, \hat{y}^{\text{upper}}_{t_k}\right]\}_{k=n+1}^{n+h}$.}
\State Compute time derivatives $\{(y'_{t_k})\}_{k=1}^{n-1}$: $y'_{t_k} \gets \frac{y_{t_{k+1}} - y_{t_k}}{t_{k+1} - t_k}$. 
\If{$\{(y_{t_k})\}_{k=1}^n$ is noisy}
    \State Smooth derivatives: $\bar{\mathbf{y}}' \gets f_{\text{smooth}}(\{y'_{t_k}\}_{k=1}^{n-1})$, where $\bar{\mathbf{y}}' = [\bar{y}'_{t_1}, \dots, \bar{y}'_{t_{n-1}}]^T$. 
    \State Derive the error due to smoothness and the sample variance as the estimator of variance:
    \[
    \boldsymbol{\delta} \gets \mathbf{y}' - \bar{\mathbf{y}}', \quad \sigma_\delta^2 \gets \frac{1}{n-2} \sum_{k=1}^{n-1} \delta_{k}^2, 
    \]
    where $\mathbf{y}' = [y'_{t_1}, \dots, y'_{t_{n-1}}]^T$, $\boldsymbol{\delta} = [\delta_{1}, \dots, \delta_{{n-1}}]^T$.
\Else
    \State Set $\bar{\mathbf{y}}' \gets \mathbf{y}', \quad \boldsymbol{\delta} \gets \mathbf{0}, \quad \sigma_\delta^2 \gets 0$. 
\EndIf
\State Construct the embedding matrix $\mathbf{X} \in \mathbb{R}^{(n-m) \times m}$ and set the response vector $\bar{\mathbf{y}}'\gets[\bar{y}'_{t_m}, \dots, \bar{y}'_{t_{n-1}}]^T$.
\State Sample $\mathbf{W}\sim p(\mathbf{W})$, $\mathbf{b}\sim p(\mathbf{b})$ such that $\mathbf{W}\in\mathbb{R}^{m\times D}, \mathbf{b}\in\mathbb{R}^D$.
\State Compute features transformed matrix $\mathbf{Z}\in\mathbb{R}^{(n-m)\times D}$, where $\mathbf{z}_i\gets\sigma(\mathbf{x}_i^T\mathbf{W}+\mathbf{b}^T)$ for $i=m, \dots, n-1$. 
\State Obtain $\boldsymbol{\theta}^{(s)}=\{\beta_0^{(s)}, \boldsymbol{\beta}^{(s)}, (\sigma_\epsilon^2)^{(s)}, (\boldsymbol{\lambda}^2)^{(s)}, (\tau^2)^{(s)}, \xi^{(s)}\}_{s=B+1}^S$ from Algorithm \ref{alg:bayesian_lasso_gibbs} with input of $\mathbf{Z}$ and $\bar{\mathbf{y}}'$, where $\boldsymbol{\lambda}^2=[\lambda_1^2, \dots, \lambda_D^2]^T$.
\For{$k=1, \dots, h$}
    \For{$s=B+1, \dots, S$}
    \State Construct an embedding matrix $\mathbf{x}^{(s)}_{n+k-1}\gets[y^{(s)}_{t_{n+k-m}}, \dots, y^{(s)}_{t_{n+k-1}}]^T$, where $y^{(s)}_{t_{n+j}}=\hat{y}^{(s)}_{t_{n+j}}$ for $j\geq1$, and $y^{(s)}_{t_{n+j}}=y_{t_{n+j}}$ otherwise.
    \State Transform feature mapping: $\mathbf{z}^{(s)}_{n+k-1}\gets\sigma((\mathbf{x}^{(s)}_{n+k-1})^T\mathbf{W}+\mathbf{b}^T)$.
    \State Predict smoothness derivative samples: 
    $\hat{\bar{y}}'^{(s)}_{t_{n+k-1}}\gets \beta_0^{(s)}+(\mathbf{z}^{(s)}_{n+k-1})^T\boldsymbol{\beta}^{(s)}+\epsilon_{n+k-1}^{(s)}$, where $\epsilon_{n+k-1}^{(s)}\sim\mathcal{N}\left(0, (\sigma_\epsilon^2)^{(s)}\right)$.
    \State Predict derivative samples: $\hat{y}'^{(s)}_{t_{n+k-1}}\gets\hat{\bar{y}}'^{(s)}_{t_{n+k-1}}+\delta_{{n+k-1}}^{(s)}$, where $\delta_{{n+k-1}}^{(s)}\sim\mathcal{N}(0, \sigma_\delta^2)$.
    \State Integrate: $\hat{y}^{(s)}_{t_{n+k}}\gets y^{(s)}_{t_{n+k-1}}+\hat{y}'^{(s)}_{t_{n+k-1}}(t_{n+k}-t_{n+k-1})$, where $y^{(s)}_{t_{n+k-1}}= y_{t_n}$ for $k=1$. 
    \EndFor
    \State $\hat{y}_{t_{n+k}}^{\text{mean}}\gets\frac{1}{S-B}\sum_{s=B+1}^S\hat{y}_{t_{n+k}}^{(s)}$.
    \State $\hat{y}_{t_{n+k}}^{\text{lower}}\gets\text{Quantile}\left(\{(\hat{y}_{t_{n+k}}^{(s)})\}_{s=B+1}^S, \frac{\alpha}{2}\right)$.
    \State $\hat{y}_{t_{n+k}}^{\text{upper}}\gets\text{Quantile}\left(\{(\hat{y}_{t_{n+k}}^{(s)})\}_{s=B+1}^S, 1-\frac{\alpha}{2}\right)$.
\EndFor    
\State \Return $\{\hat{y}^{\text{mean}}_{t_{n+1}}, \dots, \hat{y}^{\text{mean}}_{t_{n+h}}\}$, $\{\hat{y}^{\text{lower}}_{t_{n+1}}, \dots, \hat{y}^{\text{lower}}_{t_{n+h}}\}, \{\hat{y}^{\text{upper}}_{t_{n+1}}, \dots, \hat{y}^{\text{upper}}_{t_{n+h}}\}$.
\end{algorithmic}
\end{algorithm}

\section{Baseline models}\label{sec_baseline}
\subsection{Autoregressive Integrated Moving Average (ARIMA)}
Consider a stationary time series data $\{(y_{t})\}_{t=1}^n$, and define the backshift notation as $B^ky_t=y_{t-k}$, which is used to simplify the representation of lagged values \citep{robert2006time}. 
The combination of AR$(p)$ and MA$(q)$ creates ARMA$(p, q)$ model which is defined as
\begin{equation*}
    y_t=\phi_1y_{t-1}+\dots+\phi_py_{t-p}+w_t+\theta_1w_{t-1}+\dots+\theta_qw_{t-q},
\end{equation*}
where $w_t\sim \mathcal{N}(0, \sigma_w^2)$ presents white noise, and the parameters $\phi_1, \dots, \phi_p$ and $\theta_1, \dots, \theta_q$ correspond to AR$(p)$ and MA$(q)$ parts, respectively. Using backshift operators, this integration structure can be expressed as
\begin{equation*}
    (1-\phi_1B- \dots -\phi_pB^p)y_t=(1+\theta_1B+\dots+\theta_qB^q)w_t.
\end{equation*}

For non-stationary time series data, the ARMA model can be extended to form the Autoregressive Integrated Moving Average (ARIMA) model. Specifically, if $(1-B)^dy_t$ or  the $d$-th difference of the time series is stationary and can be described by an ARMA$(p, q)$ model, then a process $y_t$ is called ARIMA$(p, d, q)$ with the model written under the backshift operator as follows:
\begin{equation*}
    (1-\phi_1B-\dots-\phi_pB^p)(1-B)^dy_t=\alpha+(1+\theta_1B+\dots+\theta_qB^q)w_t,
\end{equation*}
where $w_t\sim \mathcal{N}(0, \sigma_w^2)$ is a white noise, and $\alpha=\text{E}[(1-B)^dy_t](1-\phi_1-\dots-\phi_p)$.

\subsection{Holt’s Linear Trend}
The Holt's linear trend model (Holt), or double exponential smoothing, is an expanded version of the single exponential smoothing model, by additionally modeling the trend of the time series. For further details, see 
\citet{gardner2006exponential, hyndman2018forecasting}. The model expression is
\begin{align}
\begin{aligned}
    \text{Forecast equation: } & y_{t+h}=S_t+hT_t,\\
    \text{Trend equation: } & T_t=\gamma(S_t-S_{t-1})+(1-\gamma)T_{t-1},\\
    \text{Smooth level equation: }& S_t=\alpha y_t+(1-\alpha)(S_{t-1}+T_{t-1}),
\end{aligned}
\label{holt_eq}
\end{align}
where $S_t$ is the level estimation of time series, $T_t$ is the trend estimation of the time series at time $t$, $h$ is the forecast step, $\gamma$ is the smoothing parameter of the trend, $0\leq\gamma\leq 1$, and $\alpha$ is the smoothing parameter of the level, $0\leq\alpha\leq 1$. 

\subsection{Long Short-Term Memory (LSTM)}

Long Short-Term Memory (LSTM) is a variant of Recurrent Neural Networks (RNNs) that uses special ``memory cell" units to effectively capture long-term relationships from past information without the vanishing or exploding gradient problem \citep{Graves2012}. 
Each memory cell of LSTM consists of three gates, including the forget gate, the input gate, and the output gate. The sigmoid activation function is commonly used in all three gates, with an output range of 0 to 1, which determines the percentage of information that should be kept in each gate or passed through. 
As a result, LSTMs have been widely studied in time series forecasting, as shown in \citet{fischer2018deep, siami2018comparison, fjellstrom2022long}.  


\subsection{Random Forest}

Random Forest is a well-known machine learning method introduced by \citet{breiman2001random}, which can be described as a collection of decision trees \citep{hastie2009elements}. 
In the context of classification, the final decision is determined by taking the majority vote of all trees, while in regression, the final result is the average of the predictions from all trees. 
The generalization error will converge almost surely to a limit as the number of trees increases by the Law of Large Numbers, which helps to reduce variance and improve generalization. 
Random Forests have been widely applied to time series prediction, such as forecasting COVID-19 cases and deaths \citep{ozen2024random}, stock prices \citep{qiu2017oblique, lin2017random, ghosh2022forecasting}, and the electricity market's real-time price \citep{mei2014random}.  

\section{Experiment setup}\label{sec_experiment_setup}
\subsection{Data sets}
All code and data for the experiments are available at \href{https://github.com/thuthiminhnguyen/rfBLT-numerical-experiments}{github.com/thuthiminhnguyen/rfBLT-numerical-experiments}.
\subsubsection{Simulated data}
The compartmental models are widely applied to resemble epidemiological data. The simplest model simulates the movement of people between three states, including susceptible ($S$), infected ($I$), and removed ($R$), where the removed category consists of both recovered and dead individuals. 
However, this model is too simplistic to fully capture the complex dynamics of a real pandemic. 
Several works have attempted to extend the model, including Susceptible-Exposed-Infected-Removed (SEIR) and S$\mu$EIR model, which is a variant of the SEIR model that accounts for the unconfirmed cases \citep{zou2020epidemic}. 
In this paper, the S$\mu$EIR model is used for simulation, which is expressed as follows:
\begin{align}
\begin{aligned}
    \frac{dS}{dt}&=-\frac{\beta(I+E)S}{N},\\
    \frac{dE}{dt}&=\frac{\beta(I+E)S}{N}-\sigma E,\\
    \frac{dI}{dt}&=\mu\sigma E-\gamma I,\\
    \frac{dR}{dt}&=\gamma I,
\end{aligned}
\label{SuEIR}
\end{align}
where $N=S+E+I+R$ is the population, $\beta$ is the contact rate between the susceptible and exposed groups, as well as the susceptible and infectious groups, $\sigma$ is the latency rate between the exposed and infectious groups, $\gamma$ is the transition rate between the infectious and removed groups, and $\mu$ is the discovery rate of the infected cases that are confirmed publicly.

To solve the system~\ref{SuEIR}, we use the \texttt{deSolve::ode} function in R for the time interval of $[0, 180]$ days with a day time step. The parameters outlined by \citet{saha2023spade4} are as follows:
\begin{equation*}
    \beta = 3/14, \hspace{0.5cm}\gamma=1/14, \hspace{0.5cm}\sigma=1/4, \hspace{0.5cm}\mu=3/4, 
\end{equation*}
and the initial values are
\begin{equation*}
    S_0 = 10^6, \hspace{0.5cm} E_0=0, \hspace{0.5cm} I_0=1, \hspace{0.5cm} R_0=0,\hspace{0.5cm} N=S_0+E_0+I_0+R_0.
\end{equation*}
Subsequently, we derive the proportion of cases $\{I(t)\}_{t=0}^{180}$ by taking $I/N$. In practice, the data is noisy; thus, the noise is added as follows:
\begin{equation*}
    I_{\text{noisy}}(t)=I(t)+\zeta\max_{s}|I(s)|,
\end{equation*}
where $s\in[0, 180]$, $\zeta\sim\mathcal{N}(0, \sigma_\zeta^2)$, and $\sigma_\zeta$ represents noise level.

When the data is noisy, identifying the true signal is challenging because the finite difference approach may amplify the noise. Hence, the smoothed $I_{\text{noisy}}$ is then computed, denoted as $\bar{I}_t$, by taking the left moving average with a window of 7 days, described as
\begin{equation}\label{Ismoothed}
    \bar{I}_{t}=
    \begin{cases}
        \frac{1}{t+1}\sum_{j=0}^tI_{\text{noisy}}(j), & t=0, \dots, 5,\\
        \frac{1}{7}\sum_{j=t-6}^tI_{\text{noisy}}(j), & t=6, \dots, 180.
    \end{cases}
\end{equation}

\begin{figure}[H]
  \centering
  \includegraphics[width=0.7\textwidth]{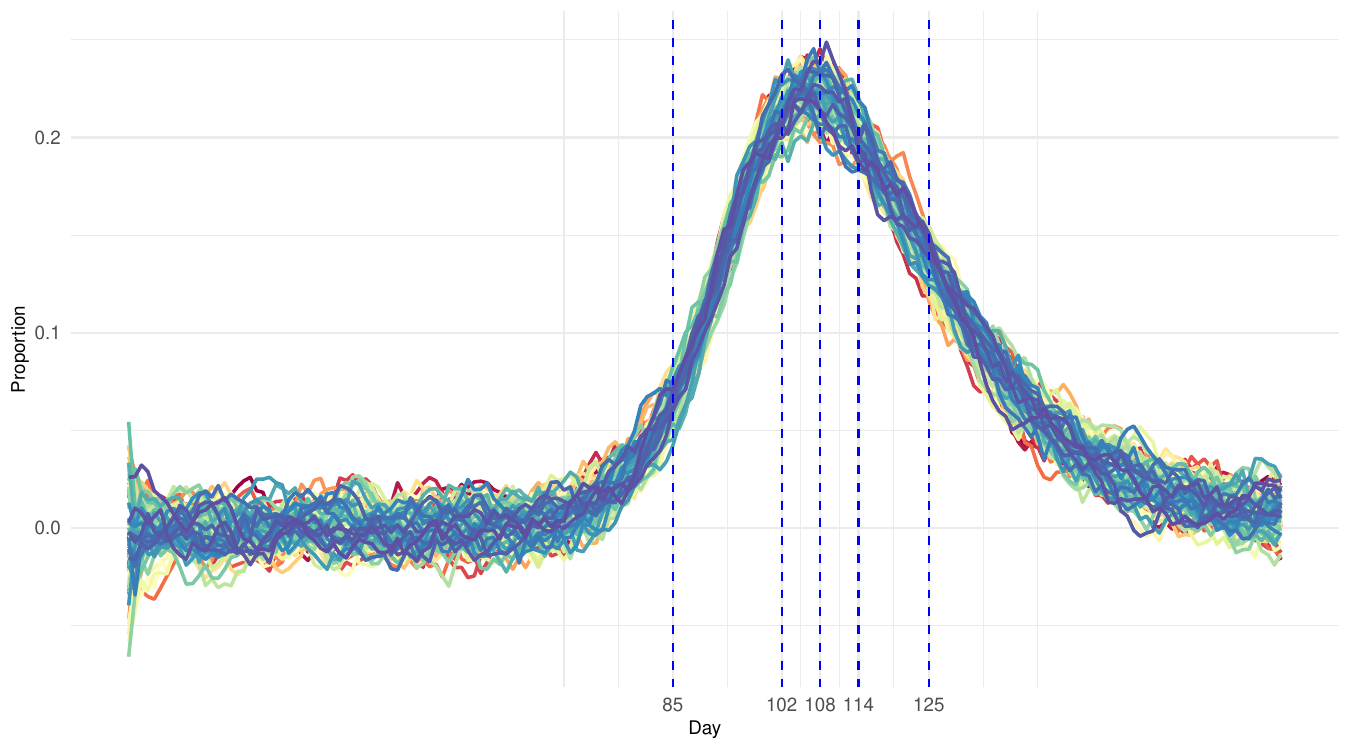}
  \caption{The figure demonstrates the 7-day left moving average of 100 infectious proportion simulations from day 0 to day 180, with a noise level of 10\%, where dashed lines represent five different starting prediction points: the 85th, 102nd, 108th, 114th, and 125th days.}
  \label{fig:simulation0.1_plot}
\end{figure}

The 100 trajectories of the smoothed infectious proportion are presented in Figure~\ref{fig:simulation0.1_plot}. Their values' range is between 0 and 0.25, rising from day 0 to a peak at about day 108, and then decreasing. Our goal is to forecast trajectories at three distinct periods: when they tend to rise (day 85), fall (days 114 and 125), and stand around the peaks (days 102 and 108).

\subsubsection{Real data}
\begin{figure}[H]
  \centering
  \begin{subfigure}[t]{0.48\textwidth}
    \centering
    \includegraphics[width=\textwidth]{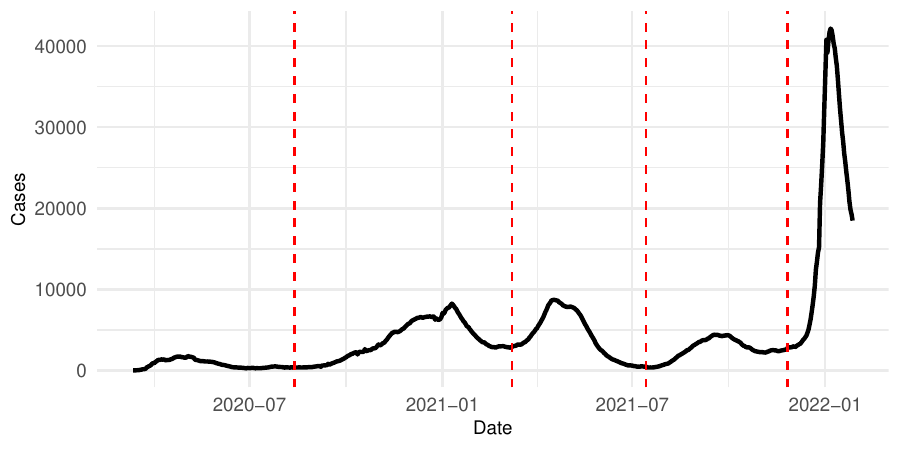}
    \caption{Daily new cases.}
    \label{fig:all_waves_new_cases_plot}
  \end{subfigure}
  \hfill
  \begin{subfigure}[t]{0.48\textwidth}
    \centering
    \includegraphics[width=\textwidth]{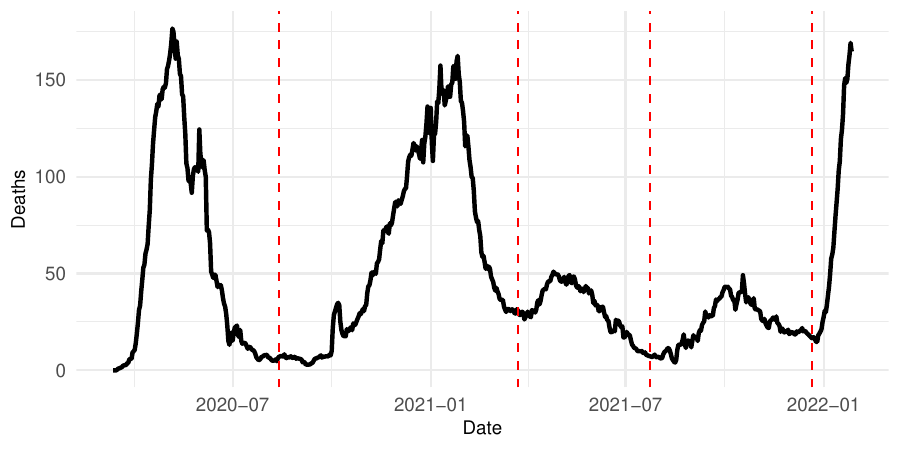}
    \caption{Daily new deaths.}
    \label{fig:all_waves_new_deaths_plot}
  \end{subfigure}
  \caption{The figures show the 7-day left-moving average of daily new COVID-19 cases (left) and deaths (right) in Canada from March 12, 2020, to January 27, 2022, with five waves separated by red dashes.}
  \label{fig:all_waves}
\end{figure}

\paragraph*{COVID-19 in Canada}
Daily COVID-19 data from March 12, 2020, to January 27, 2022, is sourced from the Government of Canada website \citep{Canada_covid}. 
During this period, there are five waves of new cases and deaths, as illustrated in Figure~\ref{fig:all_waves}, with the number of deaths coming after the number of new cases. 
Furthermore, the graphics suggest that the number of new deaths is significantly lower than the number of new cases, with new cases ranging from 0 to 40,000 and deaths from 0 to approximately 175.

In this study, we experiment with our proposed model using the second wave of the pandemic data. We employ an expanding window to forecast future values of time series in the next seven days, starting from day 101, as displayed in Figure~\ref{fig:new_cases_pred_plot} and \ref{fig:new_deaths_pred_plot} below. During this wave period, new cases ranged from 0 to 8,000, from August 13, 2020, to March 7, 2021, marked as days 1-207. Meanwhile, this wave ended on March 22 for deaths, denoted as days 1-222, with the daily new fatalities ranging from 0 to around 150. 

\begin{figure}[H]
  \centering
  \includegraphics[width=0.7\textwidth]{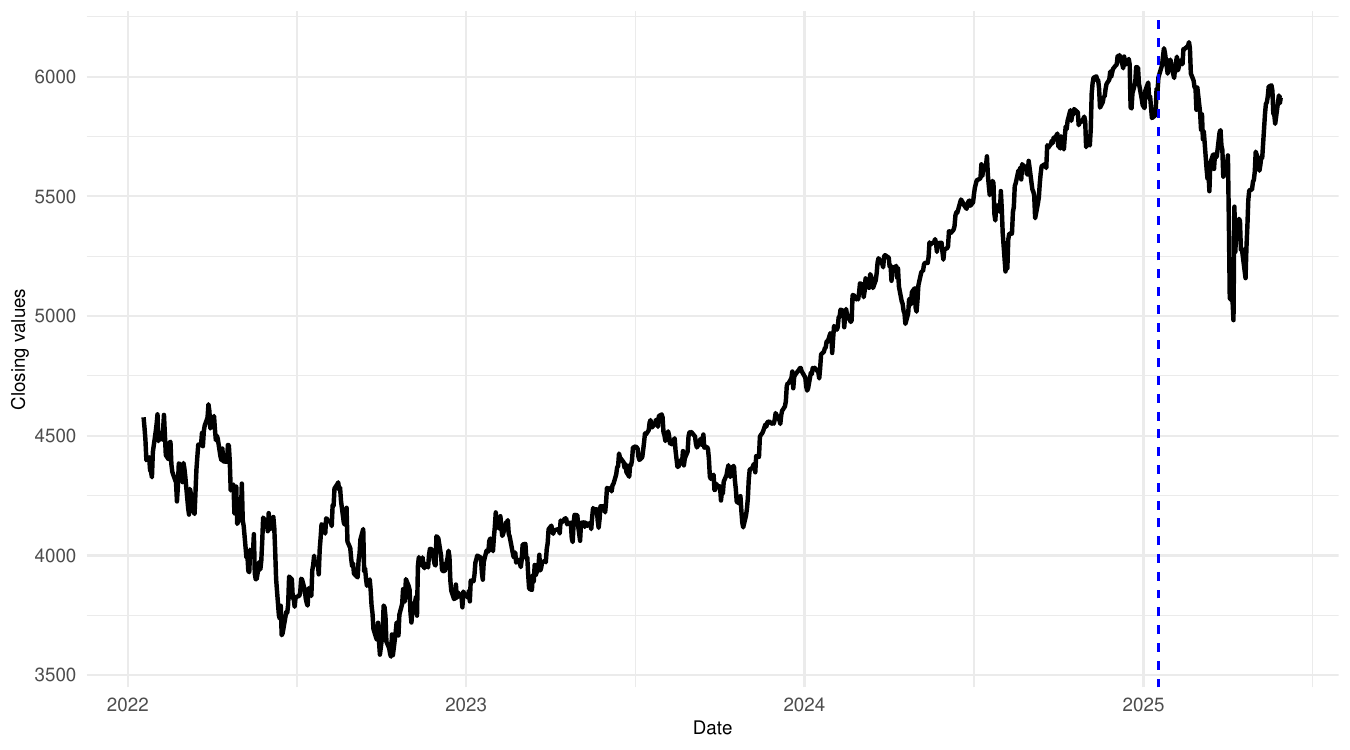}
  \caption{The figure displays daily closing values of the S\&P 500 index from January 15, 2022, to May 31, 2025, denoted as business days 1-845. The blue dashed line at the business day 753 represents the end of the first training period of the expanding window forecast.} 
  \label{fig:sp500_plot}
\end{figure}

\paragraph*{S\&P 500} The S\&P 500 index is obtained from the \texttt{quantmod} package in R \citep{quantmod2025} for the period of January 15, 2022, to May 31, 2025, represented as business days 1-845, as visualized in Figure~\ref{fig:sp500_plot}. 
We apply the expanding window 7-day forecasts starting from day 754, with the end of the initial training period marked by the blue dashed line.
The data range is from 3,500 to 6,000 with high variability. Generally, the data showed a decline from January 2022 to October 2022, followed by a rise until February 2025. Subsequently, there was a substantial drop until April 2025, and then a rebound.

\subsection{Data Preprocessing and Fitting}
\paragraph{Normalization}
In order to remove the effect of the order of magnitude in real datasets (the COVID-19 and S\&P 500), the min-max normalization is utilized, while no normalization is needed for simulated data. 
Consider time series data $\{(y_{t_k})\}_{k=1}^n$, and the forecast horizon $h$, the min-max scaling is described as follows:
\begin{equation}\label{min_max_scaling}
    \left(y_{t_k}\right)^\text{scaled} = \frac{y_{t_k}-\min(y_{t_k})}{\max(y_{t_k})-\min(y_{t_k})}.
\end{equation}

The forecast time series is then converted back to the original measurement unit, which is defined by 
\begin{equation*}
    \left(\hat{y}_{t_{n+j}}\right)^\text{converted} = \left(\hat{y}_{t_{n+j}}\right)^\text{scaled}\left[\max(y_{t_k})-\min(y_{t_k})\right]+\min(y_{t_k}),
\end{equation*}
where $j=1, \dots, h$.

\paragraph{Model Configuration}
We use an embedding window size of 9 days for all time series data, with the exception of the S\&P 500 index, which is 20 days. 
This is due to the high variability of the S\&P 500 index; therefore, the 20-day window or the previous trading month lag was chosen. 
The recursive 7-day forecast is applied to all methods. For statistical models, we report 95\% confidence and credible intervals. 
In contrast, there is no built-in confidence interval function for machine learning methods. 
The details of the models' configurations are described below.

\subparagraph{rfBLT} The model is configured with the Fourier activation function, the weighting matrix $\mathbf{W}\sim\mathcal{N}(\mathbf{0}, \mathbf{I})$, and the bias vector $\mathbf{b}\sim\text{Unif}[0, 2\pi]$. There are 2,000 samples drawn to generate posterior samples with a thinning factor of 5 and a burn-in period of 1,000 iterations. The final prediction is the mean of the posterior samples.

\subparagraph{Random Feature Bayesian Lasso (rfBL)}
The settings are the same as those of rfBLT. However, the output differs from that in Section~\ref{sec_rfBLT}; the input and output pairs are now:

\begin{equation*}
    \mathbf{X}\gets
    \begin{bmatrix}
        y_{t_1} & y_{t_2} & \dots & y_{t_{m-1}} & y_{t_m}\\
        y_{t_2} & y_{t_3} & \dots & y_{t_m} & y_{t_{m+1}}\\
        \vdots & \vdots & & \vdots & \vdots\\
        y_{t_{n-m}} & y_{t_{n-m+1}} & \dots & y_{t_{n-2}} & y_{t_{n-1}}
    \end{bmatrix}, \hspace{0.5cm}
    {\mathbf{y}}\gets\begin{bmatrix}
        y_{t_{m+1}}\\
        y_{t_{m+2}}\\
        \vdots\\
        y_{t_{n}}
    \end{bmatrix},
\end{equation*}
where $m$ is the embedding window size. 

\subparagraph{ARIMA} The model is fitted automatically. In particular, the function \texttt{forecast::auto.arima()} is applied to derive the optimal values of $p$, $d$, and $q$, using AIC as the chosen information criterion.

\subparagraph{Holt's Linear Trend method} We use the \texttt{forecast::holt()} function to forecast 7 days ahead using Equation \ref{holt_eq}. For the COVID-19 and simulated data, to account for the seasonal patterns, we use a frequency of 7 days. For the S\&P500 index, the yearly trading cycle (252 days) is applied as its frequency.

\subparagraph{LSTM} The \texttt{keras} package is used to fit LSTM models with mean squared error loss and the Adam optimizer. 
The settings are motivated by \citet{fjellstrom2022long}; however, we use the time series values as the target rather than the movement. 
This gives the input-output pairs as same as the rfBL method above. The models' configurations are varied between different datasets, which are described in Table~\ref{tab:lstm_config}. 

\begin{table}[H]
\centering
\begin{tabular}{|p{3.5cm}|>{\centering\arraybackslash}p{1.8cm}|>{\centering\arraybackslash}p{1.8cm}|>{\centering\arraybackslash}p{1.8cm}|>{\centering\arraybackslash}p{1.8cm}|}
\hline
\centering{Configuration} & {Simulations} & {New cases} & {New deaths} & {S\&P500}\\
\hline
{Hidden layers} & 3 & 3 & 3 & 2\\
{Epochs} & 200 & 300 & 300 & 200\\ 
{Hidden dimension nodes} & 64 & 64 & 64 & 64\\
{Batch size} & 64 & 64 & 32 & 32\\ 
{Learning rate} & 0.0075 & 0.0075 & 0.0075 & 0.0075\\ 
{Dropout} & 0.1 & 0.06 & 0.06 & 0.06\\ 
{Recurrent dropout} & 0.14 & 0.14 & 0.14 & 0.14\\
\hline
\end{tabular}
\caption{LSTM's configurations for simulated and real datasets.}
\label{tab:lstm_config}
\end{table}

\subparagraph{Random Forest} The \texttt{randomForest} package is utilized with the settings of \texttt{regression} type and 1,000 trees. The relative change values are fitted due to the lower relative errors; hence, the input-output pairs are
\begin{equation*}
    \mathbf{X}\gets
    \begin{bmatrix}
        r_{t_1} & r_{t_2} & \dots & r_{t_{m-1}} & r_{t_m}\\
        r_{t_2} & r_{t_3} & \dots & r_{t_m} & r_{t_{m+1}}\\
        \vdots & \vdots & & \vdots & \vdots\\
        r_{t_{n-m-1}} & r_{t_{n-m}} & \dots & r_{t_{n-3}} & r_{t_{n-2}}
    \end{bmatrix}, \hspace{0.5cm}
    {\mathbf{y}}\gets\begin{bmatrix}
        r_{t_{m+1}}\\
        r_{t_{m+2}}\\
        \vdots\\
        r_{t_{n-1}}
    \end{bmatrix},
\end{equation*}
where $m$ is the embedding window size, and the relative change $r_{t_k}=\frac{y_{t_{k+1}}}{y_{t_k}}-1$. The S\&P 500 index has $m=20$, while for the other datasets $m=9$. The predictions is then determined by $$\hat{y}_{t_{n+j}}=\hat{y}_{t_{n+j-1}}(\hat{r}_{t_{n+j-1}}+1),$$ where $j\geq 1$ and $\hat{y}_{n}=y_n$.

\subsection{Measurement Metrics}
For simulations, we generate \(N_{sim}\) trajectories, each of a fixed length \(n\), represented as \(\{y_{k,t}\}_{t=1}^n\) for \(k = 1, \ldots, N_{sim}\). We denote \(h\) as the prediction horizon and \(m\) as the specified training size of the time series such that $m\leq n-h$. For each of the \(N_{sim}\) trajectories, the training time series is $\{y_{k,t}\}_{t=1}^m$. Consequently, the prediction is $\{\hat{y}_{k,t}\}_{t=m+1}^{m+h}$, corresponding to the lower and upper bounds of confidence or credible intervals, represented as $\{\hat{y}^\text{lower}_{k,t}\}_{t=m+1}^{m+h}$ and $\{\hat{y}^\text{upper}_{k,t}\}_{t=m+1}^{m+h}$, respectively.

In the context of real datasets, a time series \(\{y_t\}_{t=1}^{n}\) is considered, along with the end of the first training period \(m\), and the prediction horizon \(h\). The expanding window forecasting method is employed with the training data \(\{y_t\}_{t=1}^v\) and the corresponding predictions \(\{\hat{y}_t\}_{t=v+1}^{v+h}\), where \(v=m, m+1, \ldots, n-h\). Consequently, the required number iterations for this process is \(n-h-m+1\). Additionally, $\{\hat{y}^\text{lower}_{t}\}_{t=v+1}^{v+h}$ and $\{\hat{y}^\text{upper}_{t}\}_{t=v+1}^{v+h}$ describe the lower and upper bounds of confidence or credible intervals. 

For representation, the following metrics are presented using the notation in the context of real-world datasets. However, they work the same for the simulations by substituting appropriate variables (e.g., the number of simulations is equivalent to the number of iterations). Moreover, the coverage probability and range are solely calculated for statistical models. 


\paragraph{Relative Error} The relative test error of each method for $h$-day ahead forecasts is determined by 
\begin{equation*}
    \text{Relative error}\left(\{y_{t}\}_{t=v+1}^{v+h}, \{\hat{y}_{t}\}_{t=v+1}^{v+h}\right)=\sqrt{\frac{\sum_{t=v+1}^{v+h}(y_{t}-\hat{y}_{t})^2}{\sum_{t=v+1}^{v+h}y_{t}^2}}.
\end{equation*}
\paragraph{Mean Directional Accuracy (MDA)} 
The directional accuracy is a binary value indicating whether the direction of the forecast time series at horizon $q=1, \dots, h$ was correct or not \citep{costantini2016forecasting} determined by 
\begin{equation*}
    \text{DA}_{{v+q}, {v}}=I\{{sgn}(y_{{v+q}}-y_{{v}})={sgn}(\hat{y}_{{v+q}}-{y}_{{v}})\},
\end{equation*}
where $I$ is the indicator. The resulting MDA of an algorithm for real data at horizon $q$ is 
\begin{equation*}
    \text{MDA}_q=\frac{1}{n-h-m+1}\sum_{v=m}^{n-h}\text{DA}_{v+q, v}.
\end{equation*}

\paragraph{Coverage Probability} The coverage probability at forecast day $q=1, \ldots, h$ is calculated as follows: 
\begin{equation*}
    \text{Coverage Probability}(q)=\frac{1}{n-h-m+1}\sum_{v=m}^{n-h}I\left\{(\hat{y}^\text{upper}_{v+q}\geq y_{v+q})\cap(y_{v+q}\geq\hat{y}^\text{lower}_{v+q})\right\}.
\end{equation*}
\paragraph{Coverage Range} The coverage range at each prediction day $q=1, \ldots, h$ is determined by 
\begin{equation*}
    \text{Coverage Range}(q)=\hat{y}^\text{upper}_{v+q}-\hat{y}^\text{lower}_{v+q}.
\end{equation*}

\section{Numerical Experiments}\label{sec_experiment_res}
\subsection*{Experiments on simulated data of 10\% noise level}
\begin{figure}[H]
  \centering
  \includegraphics[width=0.8\textwidth]{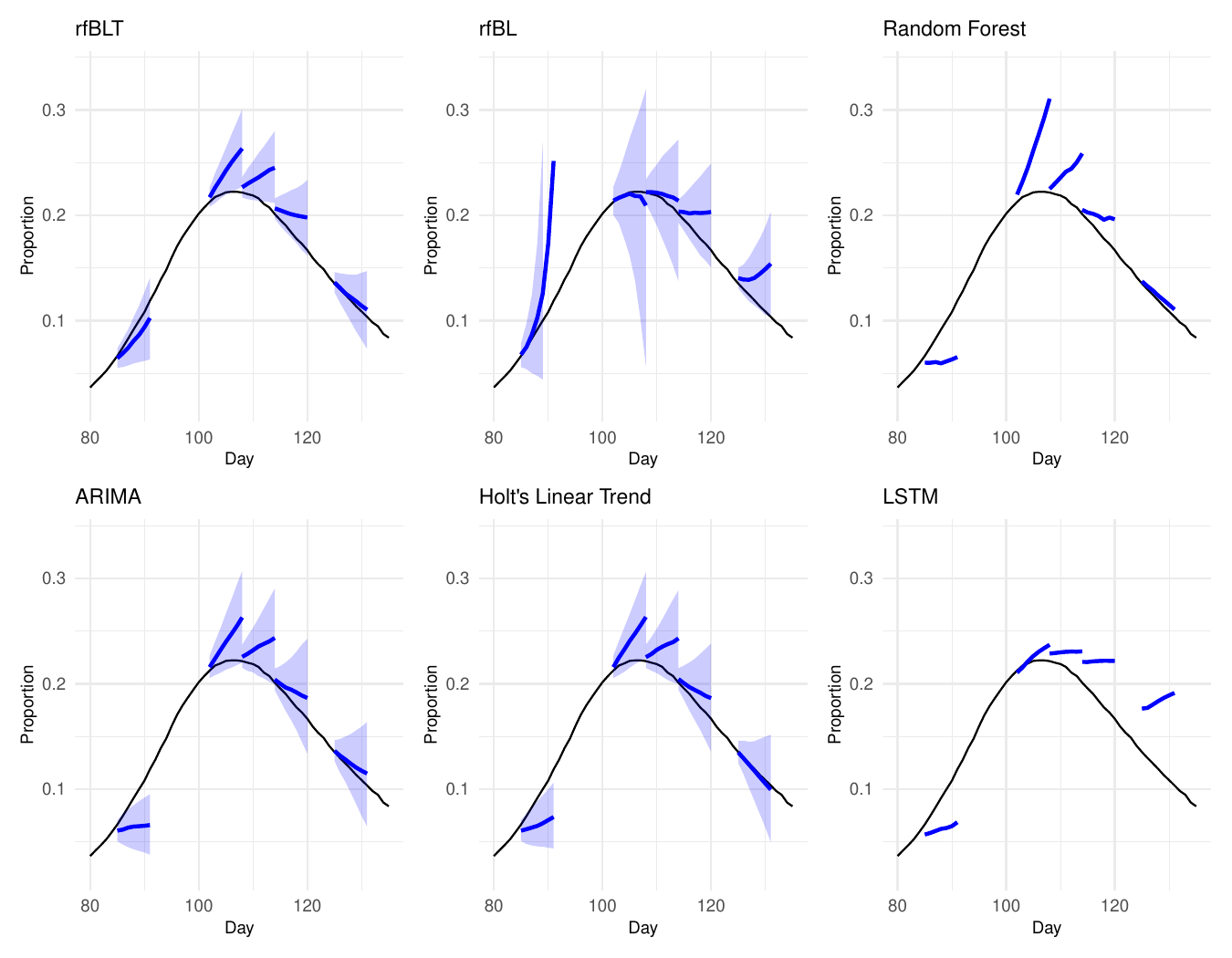}
  \caption{The figures display the median of 100 trajectories from day 80 to day 135, along with the medians of 7-day forecasts of rfBLT, rfBL, ARIMA, Holt, Random Forest, and LSTM. These forecasts are at five prediction points (days 85, 102, 108, 114, and 125). The corresponding medians of lower and upper bounds of confidence and credible intervals for statistical models are also presented.}
  \label{fig:simulation0.1_pred_plot}
\end{figure}

The first training period (days 0-84) is when the trajectories exhibit an upward trend, as displayed in Figure~\ref{fig:simulation0.1_pred_plot}. 
Most methods predict an increasing trend with a smaller slope, while rfBL generates a steeper upward prediction, failing to forecast the data movement. 
Conversely, rfBLT is the most effective model that captures the data trend by producing predictions close to the observations.

The peak occurs after the training points of days 0-101 and 0-107. 
While rfBL and LSTM capture the curvature of the data successfully, rfBLT, ARIMA, Holt, and Random Forest continue to generate upward predictions, responding to the preceding period's upward trend. 
In particular, rfBLT's predictions are similar to those of ARIMA and Holt in these two training periods. 
Although rfBLT's smoothing time derivatives help reduce variance, they also make the model less sensitive to sudden changes (at the peaks of the epidemic curves). 
By contrast, rfBL successfully captures the curvature because high-frequency fluctuations are preserved and learned by the model. 
This comes with a broad range of credible intervals, reflecting higher predictive uncertainty.

Lastly, the trajectories exhibit declining trends subsequent to the training periods of days 0-113 and days 0-124. 
For a period of 114 training days, ARIMA and Holt prove more effective than other approaches in detecting declining trends. 
Also, rfBLT and Random Forest attempt to capture it by delivering downward forecasts, but are less accurate as those of ARIMA and Holt. 
Meanwhile, rfBL gives relatively flat predictions, failing to capture the data's tendency. 
For the 125-day training period, rfBLT, ARIMA, Holt, and Random Forest outperform rfBL and LSTM in correctly recognizing the downward trend by producing forecasts close to the true values, as shown in Figure~\ref{fig:simulation0.1_pred_plot}.

\begin{figure}[H]
  \centering
  \includegraphics[width=1\textwidth]{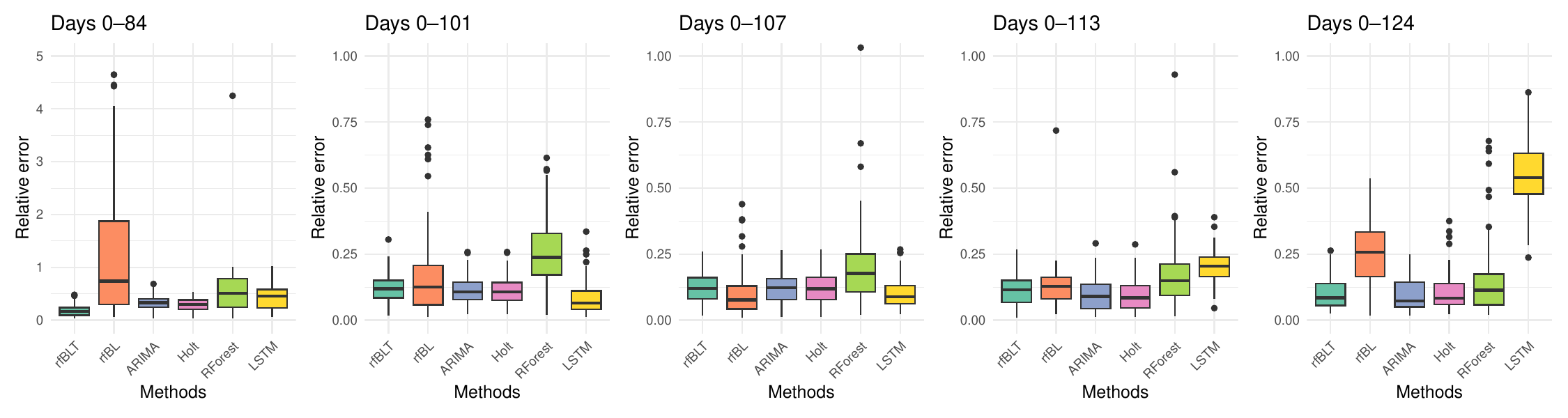}
  \caption{The figures illustrate the relative errors of 7-day forecasts from six models, based on 100 simulations conducted over five different training periods.}  
  \label{fig:simulation0.1_relative_error_all_plot}
\end{figure}

It is clear from Figure~\ref{fig:simulation0.1_relative_error_all_plot} that the relative errors of rfBLT, ARIMA, and Holt’s method 
tend to vary within a similar range and are smaller than those of the other three methods, suggesting their predictions are consistent. 
In contrast, rfBL's relative errors show high volatility at the training periods of 85, 102, and 125 days; LSTM is highly volatile 
for 125 training days, while Random Forest exhibits high volatility across all training periods.

For the training period from days 0 to 85, rfBLT produces the smallest relative error (median\(\sim0.169\)), as expected from the observed predictions, demonstrating exceptional predictive performance. 
ARIMA (median\(\sim0.333\)) and Holt (median\(\sim0.301\)) have considerably higher relative errors but also show slight variations. 
Conversely, LSTM, Random Forest, and rfBL present a remarkably wide range of relative errors, with larger medians of approximately 0.453, 0.513, and 0.741, respectively.

Prediction of peaks is challenging for most algorithms. 
For the 102-day training period, the medians' relative errors of ARIMA, Holt, rfBL, and rfBLT are approximately 0.11, roughly double that of LSTM (\(0.065\)), and about half of that of Random Forest (\(0.238\)). 
Moreover, rfBL and Random Forest exhibit an incredibly wider range of relative errors than the other four methods, indicating high volatility in their predictions.  
For the third training period (days 0-107), rfBL achieves the smallest median relative error (0.077), outperforming the other five methods. 
LSTM has a slightly higher median relative error (0.089), while the medians for rfBLT, ARIMA, and Holt are all moderately higher and roughly the same (0.12).
Meanwhile, Random Forest generates much higher relative errors, with a median of 0.177 and a wide boxplot range. 

When trajectories begin to decrease (day 114), ARIMA and Holt display the lowest median relative errors (0.087), as they successfully recognize the decreasing signal, as shown in Figure~\ref{fig:simulation0.1_pred_plot}. 
In comparison, rfBLT and rfBL have relatively higher median relative errors (0.12), while Random Forest and LSTM show substantially larger median relative errors, at nearly 0.15 and 0.2, respectively. 
Finally, all trajectories perform a downward trend after 125 days, rfBLT, ARIMA, and Holt significantly outperform the other three methods with a low median of relative errors of approximately 0.08. 
On the other hand, the median relative errors for rfBL, Random Forest, and LSTM are much higher, at approximately 0.258, 0.114, and 0.54, respectively, with significant variations.

\noindent{Note that the boxplots of relative errors are limited from 0 to 1 to improve visual clarity for training periods exceeding 100 days.}
\vspace{0.2cm}

\begin{figure}[H]
  \centering
  \begin{subfigure}[b]{1\textwidth}
      \includegraphics[width=1\textwidth]{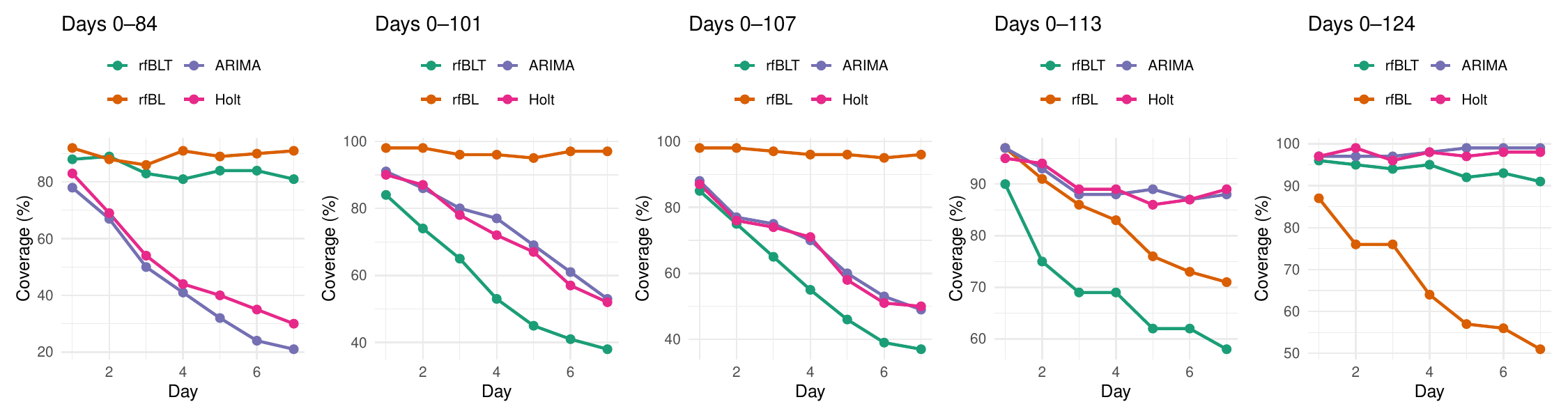}
      \caption{Coverage probability.}
      \label{fig:simulation0.1_cover_all_plot}
  \end{subfigure}\\
  \begin{subfigure}[b]{1\textwidth}
      \includegraphics[width=1\textwidth]{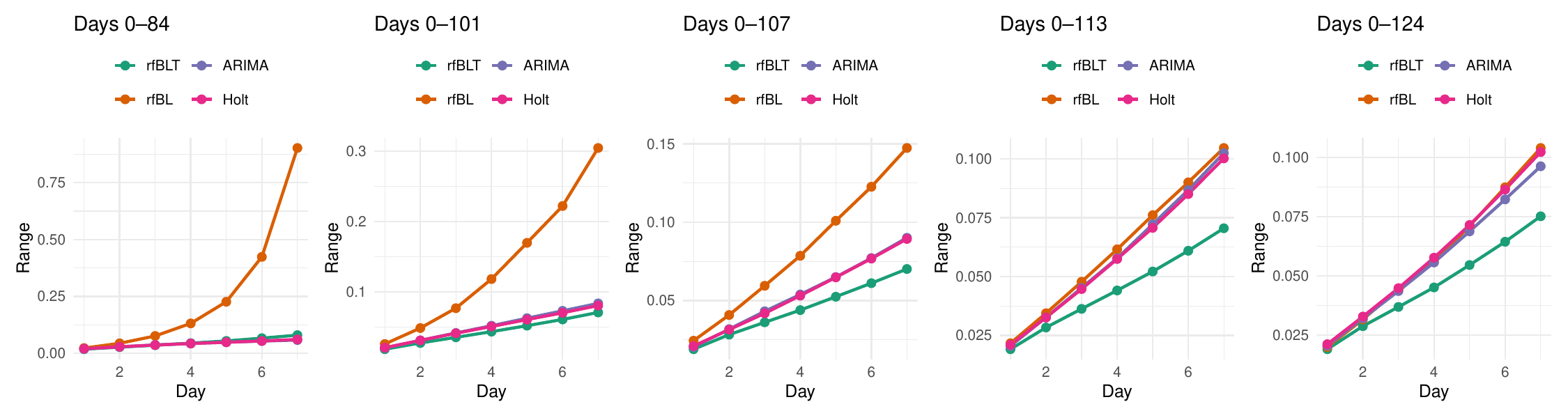}
      \caption{Median range of confidence/credible intervals.}
      \label{fig:simulation0.1_range_all_plot}
  \end{subfigure}
  \caption{The figures show the coverage percentages and median range of seven-day forecasts of four statistical models at five distinct training periods (0-84, 0-101, 0-107, 0-113, and 0-124), obtained from 100 simulations.}
  \label{fig:simulation0.1_cover_range_all_plot}
\end{figure}

Figure~\ref{fig:simulation0.1_cover_all_plot} illustrates the coverage probabilities of 7-day forecasts at five distinct training points of 100 simulations. 
In particular, they are calculated using the credible and confidence intervals obtained from statistical techniques, consisting of rfBLT, rfBL, ARIMA, and Holt's linear trend model. 
It is evident from the coverage probabilities of rfBLT that it usually works best when the trajectories are rising or falling (days 0-84 and days 0-124). 
Additionally, it tends to exhibit a smaller coverage range in comparison to ARIMA and Holt's method across all training periods, except the period of 85 days, which suggests prediction stability.
By contrast, rfBL exhibits a substantially wide range of coverages for the training periods of days 0-84, 0-101, and 0-107 as the prediction horizon increases, 
indicating high forecast uncertainty.

For the period from day 0 to day 84, the coverage probabilities of rfBLT and rfBL are relatively high, which are roughly 85\% and 90\%, respectively, 
despite rfBL's insufficient predictive performance. This is due to the wide range of credible intervals that rfBL produces as the prediction horizon increases. 
Meanwhile, the high coverage probabilities and small relative errors of rfBLT demonstrate the high reliability and precision of its forecasts.
The coverage probabilities of ARIMA and Holt, on the other hand, decline rapidly with longer prediction horizons, from roughly 80\% to about 20\% over 7 forecast days. 

At the peak (days 101 and 107), rfBL exhibits the largest coverage probabilities (around 97\%), since it not only captures the trend of trajectories 
but also produces a wide range of credible intervals. On the other hand, rfBLT, ARIMA, and Holt exhibit poor predictive performance with a narrow range 
of credible or confidence intervals. The coverage probabilities suffer as a result, dropping from roughly 90\% at the first forecast day to 40\% for 
rfBLT, and 50\% for ARIMA and Holt at the 7th prediction day. 

For the training period of days 0-113, Figure~\ref{fig:simulation0.1_cover_all_plot} shows that rfBLT's coverage probabilities continue to perform 
poorly (from about 90\% to 55\% over 7 days) as demonstrated due to its insufficient predictive performance. Meanwhile, ARIMA and Holt show a notable 
improvement, with the coverage probabilities declining from 93\% to 85\% as the prediction horizon increases. On the contrary, rfBL's coverage 
proportions show a significant drop because it generates relatively flat predictions instead of the declining forecasts. Lastly, over the training 
period of 125 days, the coverage probabilities for 7-day forecasts of rfBL drop from roughly 87\% to 50\%, because of its unsatisfactory performance 
in capturing the trend of the trajectories. In contrast, the coverage probabilities for rfBLT, ARIMA, and Holt are greatly high, approximately 
95\% for rfBLT, and 98\% for ARIMA and Holt, indicating the efficacy of their predictions.

\subsection{Experiments on daily new COVID-19 cases in Canada}
\begin{figure}[H]
  \centering
  \includegraphics[width=1\textwidth]{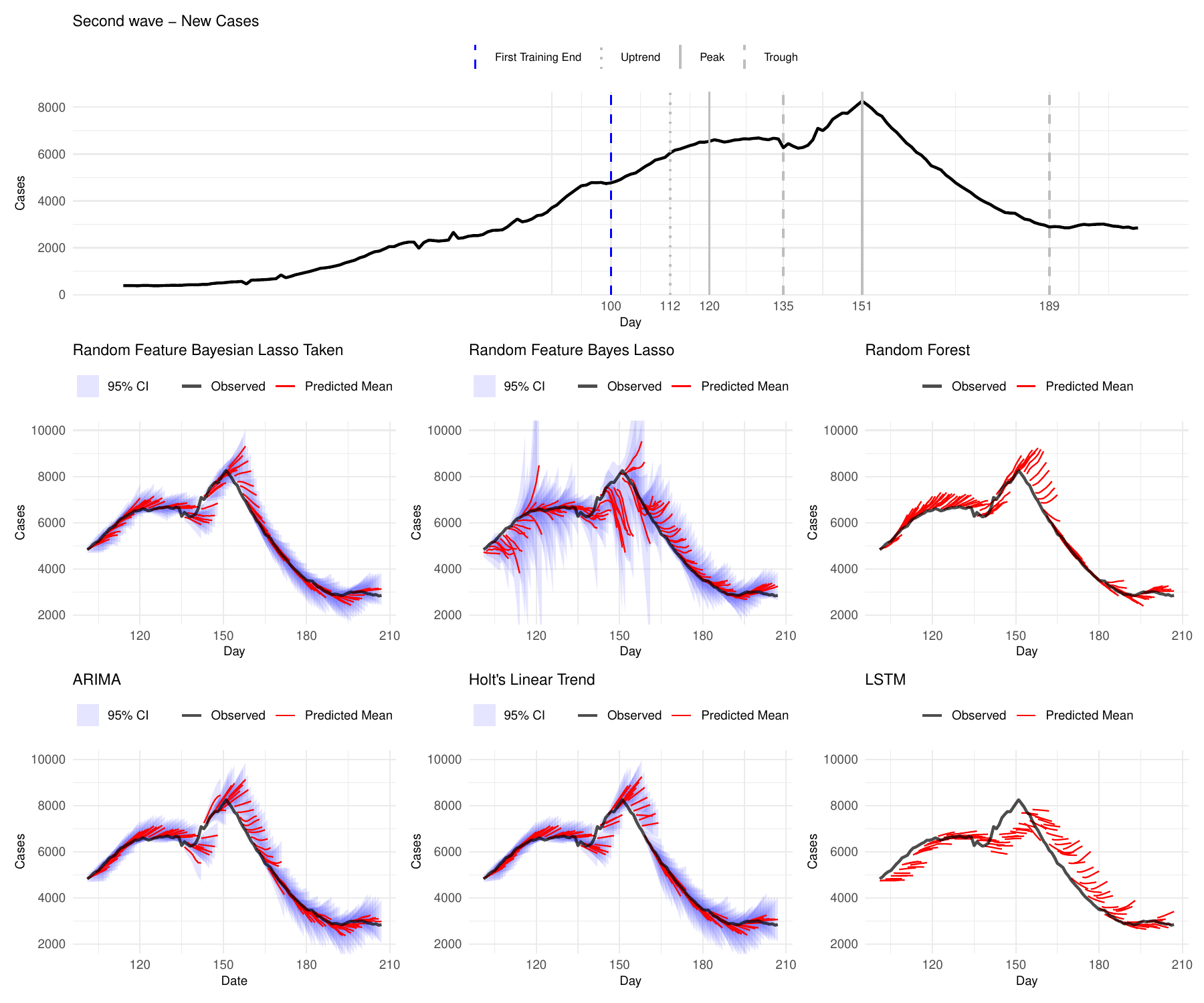}
  \caption{The first row shows the 7-day left-moving average of daily new cases trajectory, divided into subperiods for analysis. The first date of the second wave is represented as day 1, and the blue dashed line presents the end of the first training period. The next two rows present 7-day-ahead expanding window predictions starting from day 101, generated by rfBLT, rfBL, ARIMA, Holt, Random Forest, and LSTM, along with their confidence or credible intervals if applicable.}
  \label{fig:new_cases_pred_plot}
\end{figure}

The expanding window 7-day predictions of six methods beginning on the 101st day for daily new cases are displayed in 
Figure~\ref{fig:new_cases_pred_plot}. In general, the majority of methods have trouble making predictions at a trough 
(around day 135) prior to an increase in the trajectory. Additionally, all methods struggle in predicting the direction 
change from increase to decrease at the peak, around the 150th day. Among all algorithms, rfBLT's predictions 
are the closest to the true values, with reasonable credible interval ranges, demonstrating the effectiveness 
and reliability of its forecasts despite some miss forecasts of the trend at around the peak and trough.

Particularly, for the period before the 120th day, rfBLT, ARIMA, and Holt accurately predict the upward trend of the trajectory.
In contrast, rfBL and LSTM generally underpredict the true values, and Random Forest tends to overpredict during days 112-120. 
Additionally, rfBL produces significantly wider credible intervals than rfBLT, ARIMA, and Holt, suggesting greater uncertainty in 
its predictions. From days 120 to 135, when the true values are relatively flat, most methods overpredict, except rfBL and LSTM. 
At the trough that follows, the other five models (with the exception of Random Forest) produce downward predictions, suggesting they are 
trying to adapt to the sudden decrease in the time series despite the fact that the new cases increase after that. During the following period, 
the trajectory tends to rise before reaching its peak at the 151st day. rfBLT, ARIMA, Holt, and Random Forest generate 
the upward predictions that successfully capture the data movement. Conversely, rfBL and LSTM predict the opposite trend.

At around the peak, except for LSTM, the other five methods exhibit upward predictions, demonstrating responsiveness to the prior period 
when the number of new cases increases. rfBL adapts to the abrupt shift rapidly; approximately three days after the peak, it produces 
a decline forecast, but subsequently, its forecasts appear to predict the opposite direction of the actual values. Meanwhile, rfBLT adjusts a bit slower 
(about 5 days after the peak), but consistently produces decreasing forecasts later that close to the observations. This highlights the 
rapid response and predictive performance of rfBLT to the sudden change in trajectory. Meanwhile, it takes a bit longer for Random Forest, 
Holt, and ARIMA to recognize the downward trend. By contrast, LSTM generates predictions that oppose the actual tendency. 
Finally, after day 189, the trajectory slightly rises before falling. Here, the forecasts from the five methods, excluding rfBL, 
first decrease and then increase, reflecting the trend of the preceding period.

\begin{figure}[H]
  \centering
  \begin{subfigure}[b]{0.35\textwidth}
      \includegraphics[width=\textwidth]{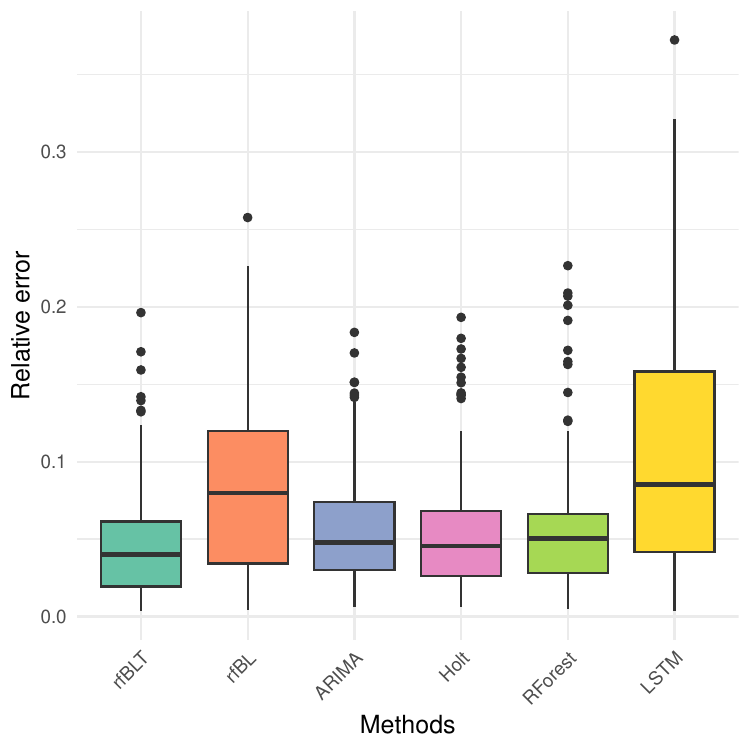}
      \caption{Relative error.}
      \label{fig:new_cases_relative_error_plot}
  \end{subfigure}
  \hspace{0.4cm}
  \begin{subfigure}[b]{0.6\textwidth}
      \raisebox{0.1cm}{\includegraphics[width=\textwidth]{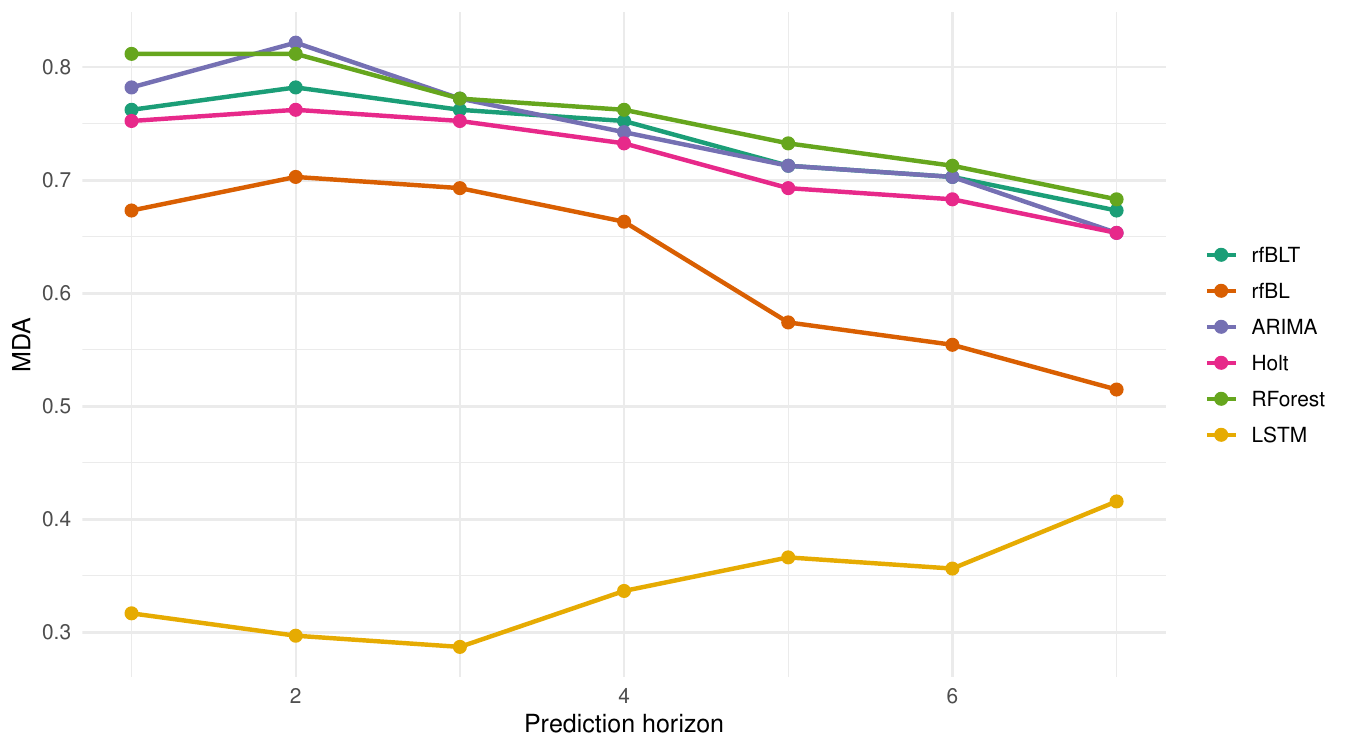}}
      \caption{Mean directional accuracy.}
      \label{fig:new_cases_MDA_plot}
  \end{subfigure}
  
  \caption{The left figure shows the relative errors between the actual and predicted values, generated by rfBLT, rfBL, ARIMA, Holt, Random Forest, and LSTM. The right figure displays the mean directional accuracy as the prediction horizon increases of all six models.}
  \label{fig:new_cases_relative_error_MDA}
\end{figure}

As shown in Figure~\ref{fig:new_cases_relative_error_plot}, rfBLT exhibits the lowest median of relative error. 
ARIMA, Holt, and Random Forest show higher medians of relative error than rfBLT, but these remain significantly lower than those of rfBL and LSTM. 
Moreover, as expected from the poor capability of predicting the trend of rfBL and LSTM, the MDA of LSTM ranges from 0.3 to 0.42, while that of rfBL lies 
between 0.5 and 0.7, as shown in Figure~\ref{fig:new_cases_MDA_plot}. 
In contrast, rfBLT, ARIMA, Holt, and Random Forest outperform rfBL and LSTM, 
even though their MDA values decline from about 0.75 to 0.65 as the prediction horizon grows.
These reinforce the conclusion about the strong predictive performance of rfBLT.

Although rfBLT has the smallest coverage ranges on five out of seven prediction days, it achieves the highest percentage of covering the true values over all the prediction time steps, as shown in Table~\ref{tab:new_cases_cover_range}. This proves that rfBLT outperforms other methods in identifying data trends and providing robust forecasts.

\begin{table}[H]
\centering
\footnotesize  
\renewcommand{\arraystretch}{1.2}
\setlength{\tabcolsep}{2.5pt}  
\begin{tabular}{|c|*{7}{c}|*{7}{c}|}
\hline
\multirow{2}{*}{Method} & 
\multicolumn{7}{c|}{Coverage Probability (\%)} & 
\multicolumn{7}{c|}{Median Range of Coverage} \\
\cline{2-15}
& Day 1 & Day 2 & Day 3 & Day 4 & Day 5 & Day 6 & Day 7
& Day 1 & Day 2 & Day 3 & Day 4 & Day 5 & Day 6 & Day 7 \\
\hline
rfBLT & \textbf{91.09} & \textbf{88.12} & \textbf{87.13} & \textbf{84.16} & \textbf{85.15} & \textbf{82.18} & \textbf{79.21} & 358.9 & 531.4 & \textbf{679.4} & \textbf{833.9} & \textbf{1004.1} & \textbf{1178.5} & \textbf{1352.2} \\
rfBL & \textbf{91.09} & \textbf{88.12} & 84.16 & 83.17 & 81.19 & 74.26 & 74.26 & 393.3 & 622.1 & 864.7 & 1156.0 & 1517.1 & 1911.4 & 2274.3 \\
ARIMA & \textbf{91.09} & 85.15 & 83.17 & 80.20 & 79.21 & 77.23 & 73.27 & \textbf{335.7} & \textbf{525.4} & 727.2 & 910.2 & 1082.9 & 1263.2 & 1475.6\\
Holt & 90.10 & 84.16 & 81.19 & 79.21 & 79.21 & 74.26 & 72.28 & 375.1 & 544.7 & 717.4 & 885.5 & 1059.4 & 1240.9 & 1428.9 \\
\hline
\end{tabular}
\caption{The table gives information about the coverage probabilities (left) and the median coverage ranges (right) across 7-day predictions of the daily new cases, generated by rfBLT, rfBL, ARIMA, and Holt. The largest value of probabilities and the smallest value of coverage range at each forecast day are highlighted in bold.}
\label{tab:new_cases_cover_range}
\end{table}

\subsection{Experiments on daily new COVID-19 deaths in Canada}
\begin{figure}[H]
  \centering
  \includegraphics[width=1\textwidth]{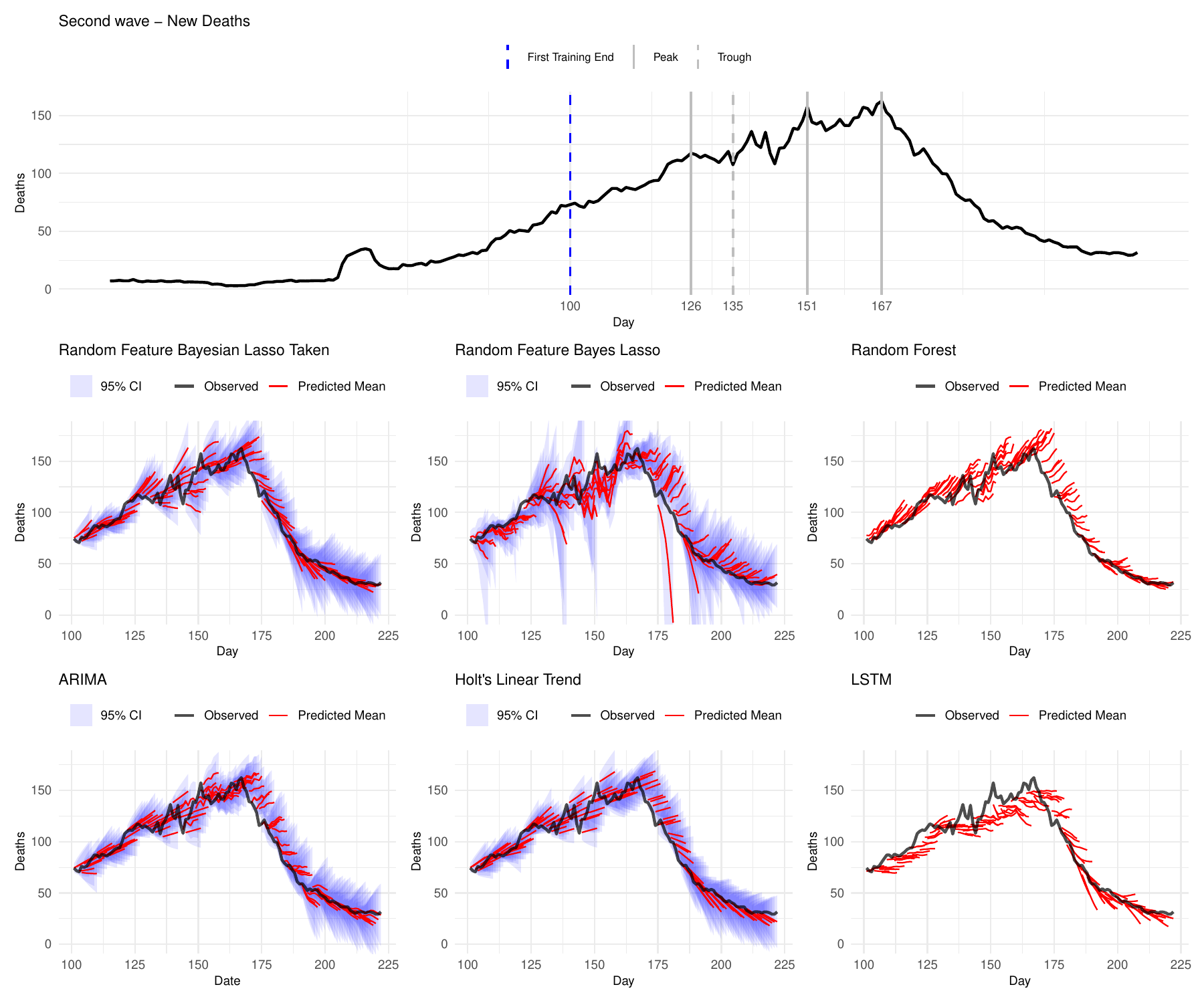}
  \caption{The first row shows the 7-day left-moving average of daily fatalities trajectory, divided into subperiods with gray lines for analysis, where the first training period starts from the 1st day to the blue dashed line. The next two rows present 7-day-ahead expanding window predictions, starting from day 101, generated by rfBLT, rfBL, ARIMA, Holt, Random Forest, and LSTM, along with their confidence or credible intervals if applicable.}
  \label{fig:new_deaths_pred_plot}
\end{figure}

The observed prediction period is highly volatile with peaks and troughs, as shown in Figure~\ref{fig:new_deaths_pred_plot}. 
Generally, before the primary peak (around day 167), most forecasting methods, excluding LSTM, consistently produce upward predictions. 
Furthermore, similar to previous sections, most algorithms have difficulty in accurately predicting the downward trend following the highest peak. 
However, rfBLT and LSTM demonstrate superior performance by quickly adapting to the immediate change from the peak.

During the period days 101-125, rfBLT, Random Forest, ARIMA, and Holt yield upward trend predictions that strongly align with the 
observations, whereas LSTM and rfBL produce relatively flat predictions, lacking of responsiveness to the data movement. 
In the subsequent period, days 126-134, except for LSTM, all other methods produce upward forecasts that are opposite to the trajectory's 
direction. From day 135 to day 151, the trajectory values are highly varied with peaks and troughs, with an increasing trend in general. 
Random Forest regularly yields upward predictions that capture the increasing trend at certain intervals, but fail to predict the decreasing 
trend at others. ARIMA, Holt, and rfBLT also generate increasing trend predictions, but with a smaller slope, while LSTM produces 
relatively flat predictions, suggesting a lack of responsiveness to the trajectory's tendency. On the other hand, rfBL produces highly 
wiggle predictions with a broad range of credible intervals in comparision to those of rfBLT, ARIMA, and Holt, suggesting rfBL's 
results are highly affected by the data's noise. During days 151-167, the data tends to decline in a small interval and then increase again. 
rfBL produces decreasing and then increasing forecasts with high variability. While rfBLT, ARIMA, Holt, and Random Forest give upward 
forecasts that fail to capture the abrupt decrease, they still recognize the overall increase trend during this period. On the other hand, 
LSTM continuously generates downward predictions. 

At the highest peak (day 167), except for LSTM, all algorithms continue to give upward predictions, failing to forecast the decrease change. 
After the peak, the trajectory's values start to drop rapidly. rfBLT begins to produce decreasing predictions after about three days of the peak; 
likewise, LSTM produces downward forecasts after about seven days of the peak. On the contrary, rfBL, Random Forest, ARIMA, and Holt's linear 
trend yield upward predictions up to around day 180, and then change to decreasing predictions that align with the data tendency. 
This demonstrates the remarkable and quick adaptation ability of rfBLT to changes in the time series.

\begin{figure}[H]
  \centering
  \begin{subfigure}[b]{0.35\textwidth}
      \includegraphics[width=\textwidth]{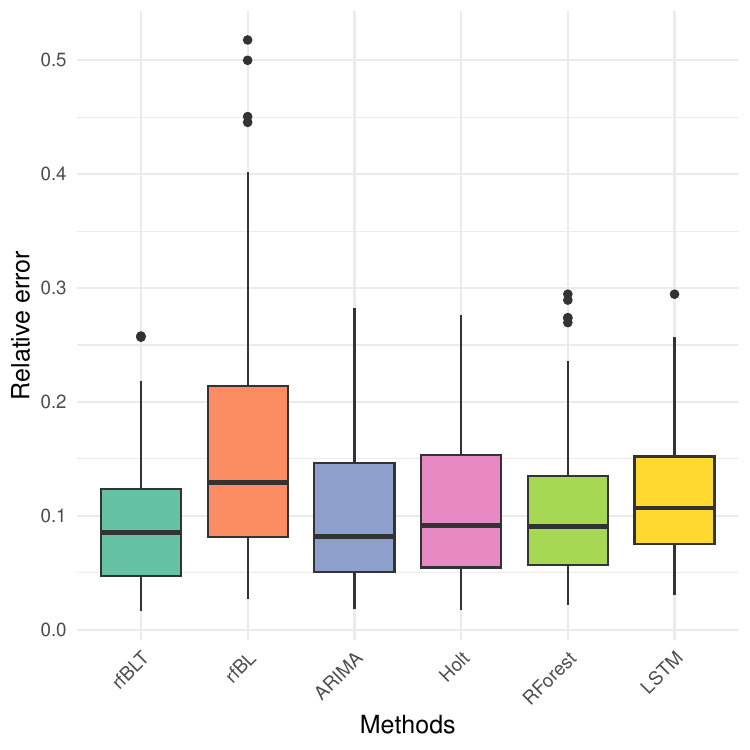}
      \caption{Relative error.}
      \label{fig:new_deaths_relative_error_plot}
  \end{subfigure}
  \hspace{0.4cm}
  \begin{subfigure}[b]{0.6\textwidth}
      \raisebox{0.1cm}{\includegraphics[width=\textwidth]{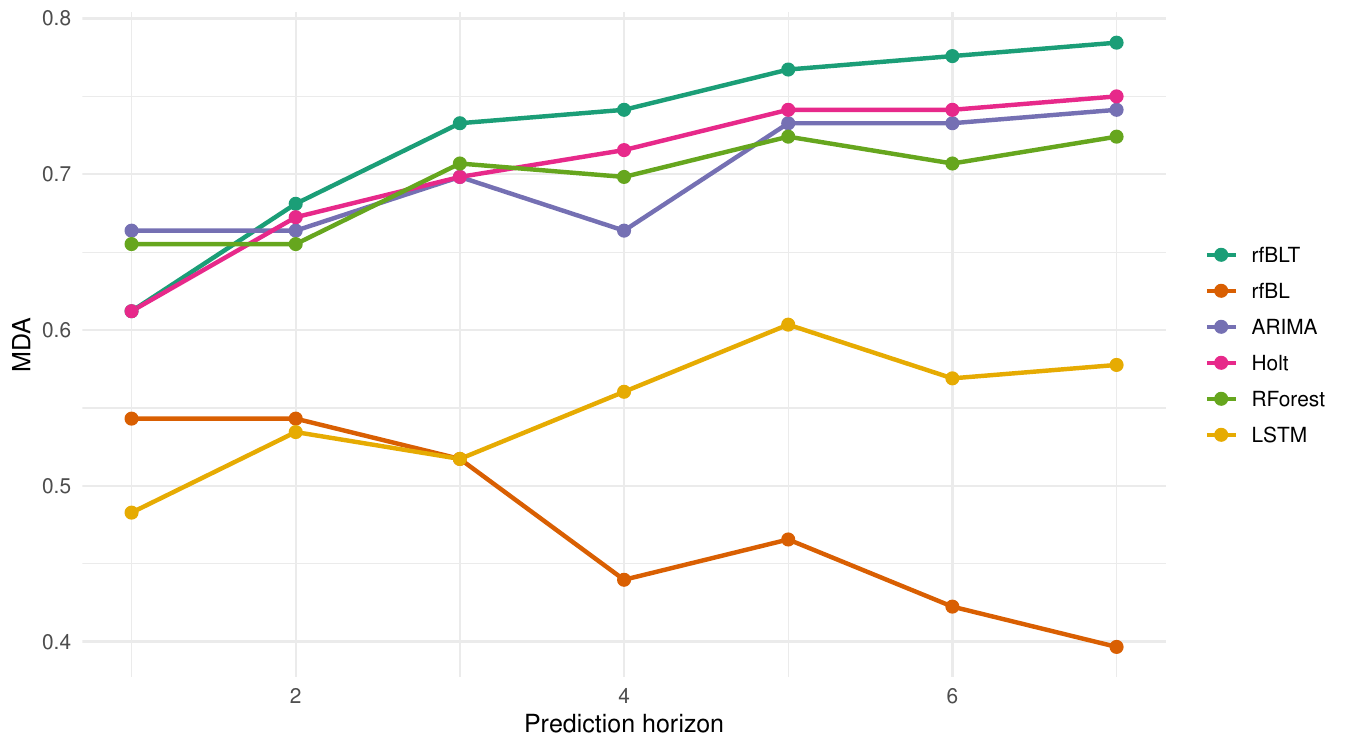}}
      \caption{Mean directional accuracy.}
      \label{fig:new_deaths_MDA_plot}
  \end{subfigure}
  \caption{The figures show the relative errors (left) and mean directional accuracy (right) between the actual and predicted values of daily new deaths as the prediction horizon grows.}
  \label{fig:new_deaths_relative_error_MDA}
\end{figure}

Figure~\ref{fig:new_deaths_relative_error_plot} shows that the median relative errors of ARIMA is the smallest, followed by rfBLT with a 
comparable level but less variation. Holt and Random Forest have slightly higher medians of relative errors compared to rfBLT. The medians 
of rfBL and LSTM's relative errors, on the other hand, are significantly higher than those of rfBLT and vary greatly (particularly rfBL). 
In terms of MDA, Figure~\ref{fig:new_deaths_MDA_plot} suggests that rfBLT substantially outperforms the rest of the other methods as the 
prediction horizon rises, increasing from around 0.6 to nearly 0.8. Meanwhile, the MDAs of ARIMA, Holt, and Random Forest exhibit similar 
patterns over the forecast period, ranging from approximately 0.65 to 0.75, while those of rfBL and LSTM remain under 0.6.
These results suggest the impressive performance of rfBLT in both precision and trend projection.

Table~\ref{tab:new_deaths_cover_range} shows that ARIMA outperforms other methods by producing confidence intervals that most frequently 
cover the true values with smaller ranges compared to other methods. rfBLT generates comparable coverage probabilities; however, 
the credible intervals of rfBLT widen as the prediction horizon grows, indicating greater uncertainty in its predictions for this noisy data. 

\begin{table}[H]
\centering
\footnotesize  
\renewcommand{\arraystretch}{1.2}
\setlength{\tabcolsep}{2.8pt}  
\begin{tabular}{|c|*{7}{c}|*{7}{c}|}
\hline
\multirow{2}{*}{Method} & 
\multicolumn{7}{c|}{Coverage Probability (\%)} & 
\multicolumn{7}{c|}{Median Range of Coverage} \\
\cline{2-15}
& Day 1 & Day 2 & Day 3 & Day 4 & Day 5 & Day 6 & Day 7
& Day 1 & Day 2 & Day 3 & Day 4 & Day 5 & Day 6 & Day 7 \\
\hline
rfBLT & 81.90 & 81.90 & \textbf{80.17} & \textbf{81.90} & \textbf{83.62} & 79.31 & 79.31 & 15.7 & 22.9 & 28.2 & 33.4 & 38.6 & 43.4 & 48.2 \\
rfBL & 81.03 & 74.14 & 75.00 & 73.28 & 69.83 & 68.10 & 70.69 & 15.1 & 23.1 & 29.2 & 36.5 & 44.5 & 51.6 & 59.0 \\
ARIMA & 81.90 & \textbf{82.76} & \textbf{80.17} & \textbf{81.90} & 80.17 & \textbf{81.90} & \textbf{81.03} & \textbf{14.3} & \textbf{21.0} & \textbf{24.4} & \textbf{28.9} & \textbf{32.4} & \textbf{35.0} & \textbf{38.1}\\
Holt & \textbf{85.34} & 80.17 & \textbf{80.17} & 80.17 & 77.59 & 79.31 & 76.72 & 16.3 & 21.9 & 26.4 & 30.2 & 33.5 & 36.7 & 39.6 \\
\hline
\end{tabular}
\caption{The table summarizes the coverage probabilities (left) and the median coverage ranges (right) across 7-day predictions of the daily deaths, generated by rfBLT, rfBL, ARIMA, and Holt. The largest value of probabilities and the smallest value of coverage range at each forecast day are highlighted in bold.}
\label{tab:new_deaths_cover_range}
\end{table}

\subsection{Experiments on S\&P 500}
\begin{figure}[H]
  \centering
  \includegraphics[width=1\textwidth]{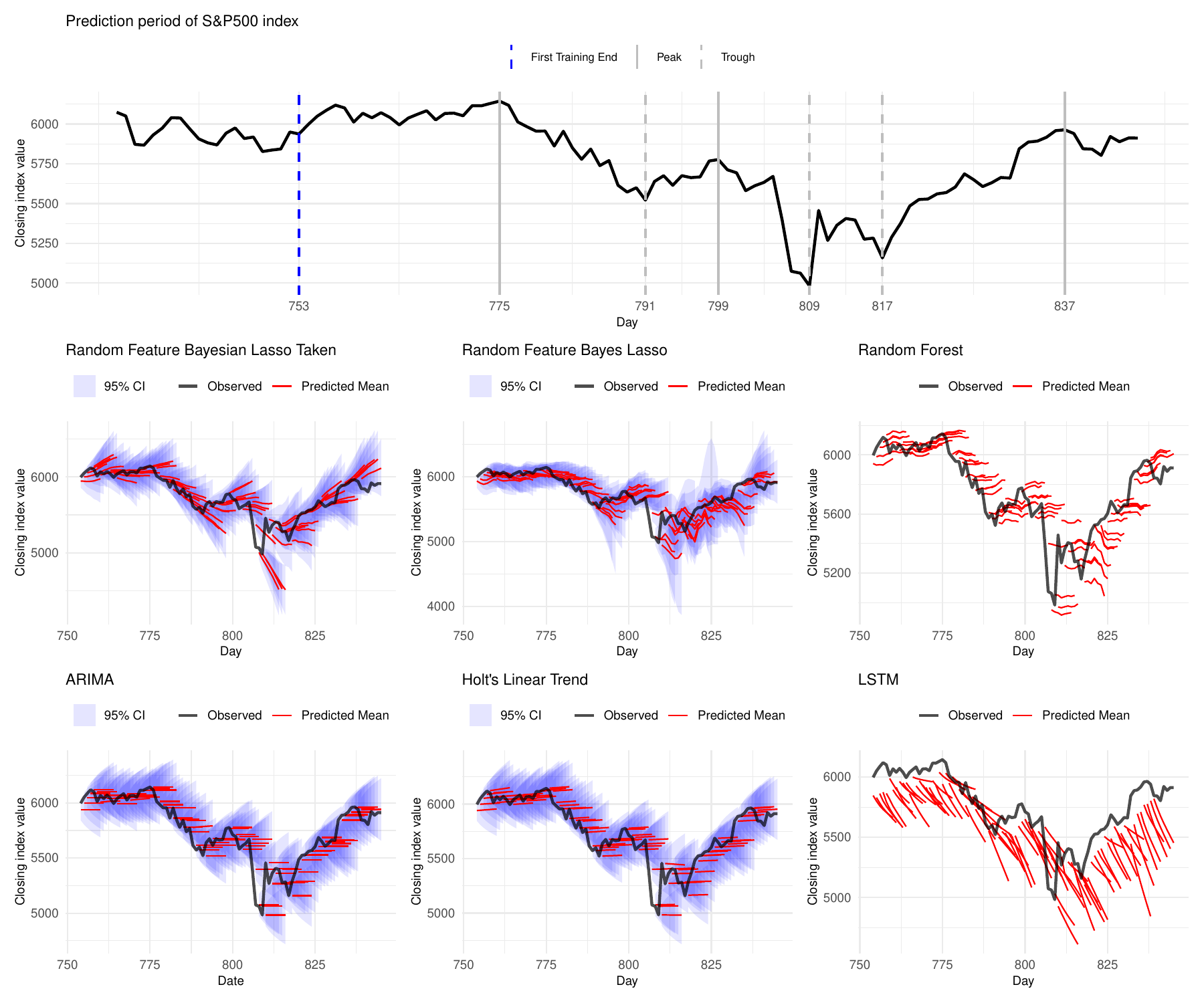}
  \caption{The first row shows the daily S\&P 500 closed index starting from the 733rd day. The trajectory is divided into subperiods with gray lines for analysis, where the first training period spans from the 1st day to the 753rd day (blue dashed line). The next two rows present 7-day-ahead expanding window predictions, starting from day 754, generated by rfBLT, rfBL, ARIMA, Holt, Random Forest, and LSTM, along with their confidence or credible intervals if applicable.}
  \label{fig:sp500_pred_plot}
\end{figure}

The prediction period can be divided into several small intervals for detailed analysis, as demonstrated in Figure~\ref{fig:sp500_pred_plot}. 
In general, Random Forest provides flat forecasts with slight fluctuations, ARIMA and Holt produce linearly flat predictions over all 
subintervals, while LSTM continuously gives decrease forecasts. While rfBL's predictions appear to somewhat match the data movement 
in the first half of the evaluated period, they fluctuate dramatically in the second half as a result of taking into account data noise. 
On the other hand, rfBLT's forecasts consistently and strongly align with the data's trend, whether it is upward, downward, or flat, 
showing the model's effectiveness in projecting the trajectory across different periods. Notably, all four statistical methods 
generate a wide range of coverage, reflecting uncertainty in predicting this highly volatile trajectory. 

During the period of days 754-775, the trajectory's values slightly fluctuate. rfBLT produces fairly smooth predictions, 
except at some small peaks, which give gently upward forecasts. rfBL and Random Forest also yield relatively flat predictions with minor fluctuations. 
In the next period, from day 776 to 791, the trajectory declines gradually. 
ARIMA, Holt, rfBL, and Random Forest all generate horizontal forecasts with minor fluctuations, failing to identify the decrease trend.
Conversely, rfBLT successfully yields declining predictions. During days 792-799, the time series slightly increases after the trough. 
It is evident that predictions from most methods, except rfBLT and rfBL, do not align with the change of the trajectory's direction. 
It takes about 4 days for rfBLT and rfBL to adapt to the change and generate upward predictions, indicating their rapid adjustment of to the sudden change. 

In the period, from day 800 to 817, the trajectory fluctuates notably with a sharp drop at day 809. 
Excluding LSTM, the other methods struggle to adjust and produce decreasing projections.
rfBLT's forecasts tend to decrease after about 3 days, but with a less steep slope than the actual series, and continue to decline more steeply at the trough, failing to project the upward trend.
Meanwhile, rfBL gives downward jiggle predictions after the significant drop (days 809-816). 
The following period is from day 817 to day 837, during which the time series shows an increase. 
rfBLT successfully derives increased predictions, while the other five methods fail to recognize the new trend. 
Lastly, beginning from day 837, the trajectory exhibits a slight decrease; except for rfBL, the other five methods fail to detect this change.

\begin{figure}[H]
  \centering
  \begin{subfigure}[b]{0.35\textwidth}
      \includegraphics[width=\textwidth]{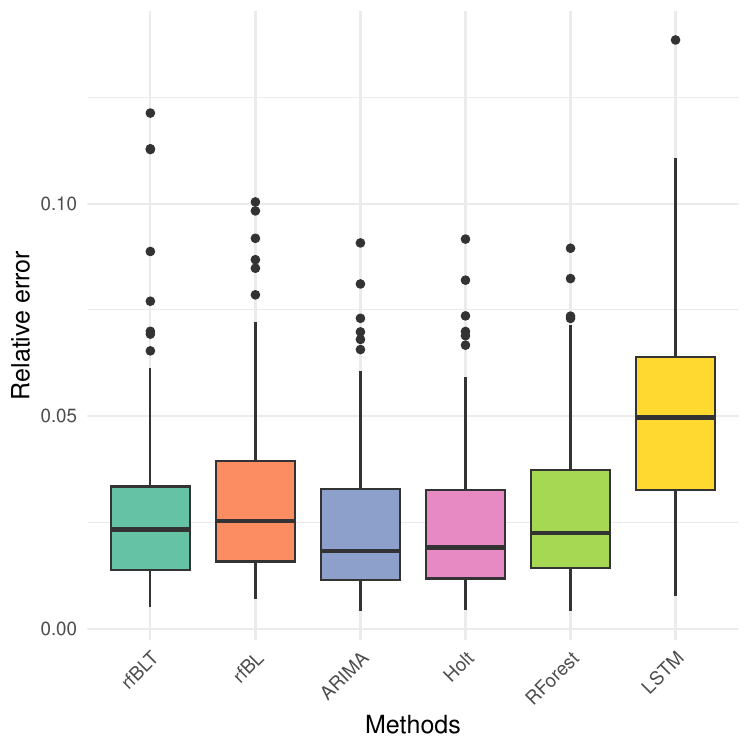}
      \caption{Relative error.}
      \label{fig:sp500_relative_error_plot}
  \end{subfigure}
  \hspace{0.4cm}
  \raisebox{0.2cm}{ 
  \begin{subfigure}[b]{0.6\textwidth}
      \raisebox{0.1cm}{\includegraphics[width=\textwidth]{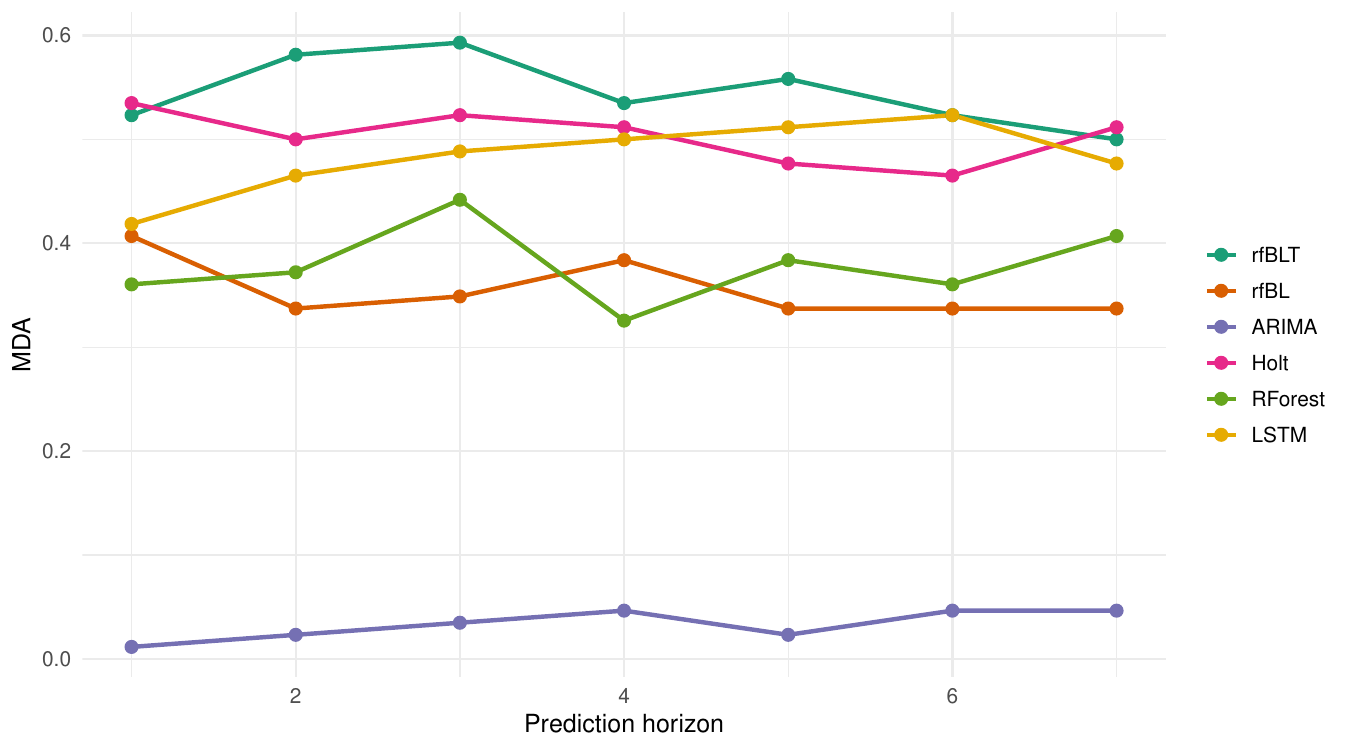}}
      \caption{Mean directional accuracy.}
      \label{fig:sp500_MDA_plot}
  \end{subfigure}}
  \caption{The figures show the relative errors (left) and mean directional accuracy (right) between the actual and predicted values of the S\&P 500 index as the forecast horizon grows.}
  \label{fig:sp500_relative_error_MDA}
\end{figure}

Figure~\ref{fig:sp500_relative_error_plot} shows that ARIMA and Holt achieve the lowest median errors. 
Following them, Random Forest and rfBLT have slightly higher median errors, rfBL's median relative error is moderately higher, and LSTM produces significantly higher median relative errors than the other methods. 
In terms of MDA, ARIMA presents the lowest MDAs, ranging from approximately 0.03 to 0.05 at all prediction horizons, despite its low median relative errors, as illustrated in Figure~\ref{fig:sp500_MDA_plot}. 
This occurs because ARIMA yields flat predictions over the forecast horizon, failing to identify trends.
By contrast, rfBLT has a slightly higher median relative error but delivers the highest MDAs (around 0.5-0.6), outperforming the other five methods.
In addition, while Holt's linear model produces predictions with similar patterns to those of ARIMA, its MDAs are slightly smaller than those of rfBLT, ranging from approximately 0.43 to 0.53.
Also, despite high relative errors, LSTM's MDAs are comparable to those of Holt. These occur because MDA does not account for the magnitude of the forecasts. 
Finally, rfBL and Random Forest achieve intermediate MDAs, higher than ARIMA, but lower than rfBLT, LSTM, and Holt, falling between 0.33 and 0.43.
These results demonstrate the effectiveness of rfBLT in capturing the trajectory's trend and producing precise forecasts.

Table~\ref{tab:sp500_cover_range} summarizes the coverage probabilities and ranges for a 7-day forecasting period using rfBLT, rfBL, ARIMA, and Holt. 
rfBLT performs poorly during the first three prediction days but significantly outperforms the other three methods for the rest of the prediction horizon. 
This happens due to the small range of credible intervals in the initial three prediction days compared to the dominant methods on those specific days. 
Furthermore, we observe that as the forecast time step increases, the median coverage ranges also increase.
These results show that rfBLT attempts to account for the uncertainty of the high-volatility trajectory as the forecast horizon rises.

\begin{table}[H]
\centering
\footnotesize  
\renewcommand{\arraystretch}{1.2}
\setlength{\tabcolsep}{2.8pt}  
\begin{tabular}{|c|*{7}{c}|*{7}{c}|}
\hline
\multirow{2}{*}{Method} & 
\multicolumn{7}{c|}{Coverage Probability (\%)} & 
\multicolumn{7}{c|}{Median Range of Coverage} \\
\cline{2-15}
& Day 1 & Day 2 & Day 3 & Day 4 & Day 5 & Day 6 & Day 7
& Day 1 & Day 2 & Day 3 & Day 4 & Day 5 & Day 6 & Day 7 \\
\hline
rfBLT & 76.74 & 76.74 & 76.74 & \textbf{82.56} & \textbf{84.88} & \textbf{86.05} & \textbf{87.21} & \textbf{181.7} & \textbf{269.2} & 349.8 & 423.6 & 492.0 & 568.3 & 644.2\\
rfBL & \textbf{86.05} & 83.72 & \textbf{82.56} & 81.40 & 75.58 & 79.07 & 77.91 & 214.2 & 293.8 & 369.8 & 429.6 & 491.7 & 534.8 & 570.7\\
ARIMA & 80.23 & \textbf{84.88} & 80.23 & 81.40 & 79.07 & 79.07 & 76.74 & 191.7 & 271.1 & 332.0 & 383.4 & 428.6 & 469.5 & \textbf{507.1}\\
Holt & 80.23 & 83.72 & 77.91 & 81.40 & 76.74 & 79.07 & 76.74 & 192.1 & 270.9 & \textbf{331.6} & \textbf{383.0} & \textbf{428.4} & \textbf{469.3} & 507.7\\
\hline
\end{tabular}
\caption{The table summarizes the coverage probabilities (left) and the median coverage ranges (right) across 7-day predictions of the daily S\&P500 closed index, generated by rfBLT, rfBL, ARIMA, and Holt. The largest value of probabilities and the smallest value of coverage range at each forecast day are highlighted in bold.}
\label{tab:sp500_cover_range}
\end{table}

\section{Conclusion and Discussion}\label{sec_conclusion_discuss}
In this paper, we present the rfBLT algorithm to forecast future non-periodic time series, available via the \texttt{rfBLT} package. The method leverages random 
feature mappings to address non-linear relationships in time delay embeddings, while employing regularized regression within the Bayesian framework to obtain 
key coefficients and prevent overfitting. We compare rfBLT with baseline models, including statistical models (ARIMA and Holt's linear trend), and machine learning 
models (LSTM and Random Forest). The outcomes from simulated and real data demonstrate the exceptional effectiveness of rfBLT in forecasting the tendency of trajectories, 
supported by the results of MDA and coverage probability. In particular, for high-volatile time series, rfBLT outperforms ARIMA in directional accuracy, as evidenced by superior MDA results.
Additionally, rfBLT's relative errors are not only comparable to those of the best methods, but they also exhibit less volatility, especially compared to the rfBL model. 
This demonstrates high precision and solidifies its efficacy in using the time-delayed embedding to reconstruct dynamical systems.
Moreover, the ability to measure prediction uncertainty by providing credible intervals, achieved by continuously updating the posterior distribution, is the key advantage of rfBLT over complex machine learning models. 

rfBLT's performance is influenced by a variety of factors that can be further studied. First, the target variable of rfBLT is the smooth time derivatives, meaning the filtering function may partially 
affect the final result. Although a simple 7-day left-moving average is applied in this study, other denoising techniques might be needed for data with a high noise level. 
Second, the selection of the number of random features should balance between computational cost and model complexity. Here, we set approximately half of the training 
data points as the number of features; however, more features might be required to capture the underlying nonlinear relationships. Third, the distribution 
options of the weighting matrix and biases and the activation function are dependent on whether the data is seasonal or non-seasonal; hence, the extension of rfBLT for forecasting periodic data is a direction for future work. 
Another hyperparameter is the embedding window size. By Takens' theorem, the embedding dimension of a deterministic system should be at least $2d+1$, where $d$ is the number of variables in the system. 
However, not all time series can be represented by a deterministic dynamics. Further research could explore the optimal embedding size using the information criterion. 

\section*{Declaration of generative AI in the writing process}
This work was prepared with the assistance of Gemini of Google AI to enhance language and readability. 
We have reviewed and edited all of its contributions and take full responsibility for the content of the published article.

\section*{Acknowledgements}
LSTH was supported by the Canada Research Chairs program and the NSERC Discovery Grant.

\appendix
\section{Gibbs Sampling for Bayesian Ridge with Gaussian Errors}
\setcounter{algorithm}{0}
The below algorithm describes Gibbs Sampling for Ridge regression with Gaussian errors under Bayesian framework, as studied by \citet{makalic2016high}. 
\label{app:bayesian_ridge_gibbs}
\begin{algorithm}[H]
\caption{Gibbs Sampling for Bayesian Ridge with Gaussian Errors}
\label{alg:bayesian_ridge_gibbs}
\begin{algorithmic}[1]
\State \textbf{Input}: Data \(\mathbf{y} \in \mathbb{R}^n\), predictor matrix \(\mathbf{Z} \in \mathbb{R}^{n \times D}\), number of samples \(S\), burn-in \(B\).
\State \textbf{Output}: Posterior samples \(\{\beta_0^{(s)}, \boldsymbol{\beta}^{(s)}, (\sigma_\epsilon^2)^{(s)}, (\tau^2)^{(s)}, \xi^{(s)}\}_{s=B+1}^S\).
\State \textbf{Initialize}:
\State \quad \(\beta_0^{(0)} \), \(\boldsymbol{\beta}^{(0)}\), \((\sigma_\epsilon^2)^{(0)}\), \((\tau^2)^{(0)}\), \(\xi^{(0)}\).
\For{\(s = 1, \ldots, S\)}
    \State \textbf{Sample \(\beta_0^{(s)}\)}:
    \[
    \beta_0^{(s)} \sim \mathcal{N}\left( \frac{1}{n} \sum_{i=1}^n \left( y_i - \mathbf{z}_i^T \boldsymbol{\beta}^{(s-1)} \right), \frac{(\sigma_\epsilon^2)^{(s-1)}}{n} \right).
    \]
    \State \textbf{Sample \(\boldsymbol{\beta}^{(s)}\)}:
    \[
    \boldsymbol{\beta}^{(s)} \sim \mathcal{N}_D\left( \tilde{\mu}, \tilde{\boldsymbol{\Lambda}}_D^{-1} \right),
    \]
    \[
    \tilde{\boldsymbol{\Lambda}}_D = \frac{1}{(\sigma_\epsilon^2)^{(s-1)}} \mathbf{Z}^T \mathbf{Z} + \frac{1}{(\sigma_\epsilon^2)^{(s-1)} (\tau^2)^{(s-1)}}\mathbf{I}_D,
    \]
    \[
    \tilde{\mu} = \tilde{\boldsymbol{\Lambda}}_D^{-1} \left( \frac{1}{(\sigma_\epsilon^2)^{(s-1)}} \mathbf{Z}^T \left( \mathbf{y} - \beta_0^{(s)} \mathbf{1}_n \right) \right).
    \]
    \Comment{Use \citet{rue2001fast} if \(D/n < 2\), \citet{bhattacharya2016fast} if \(D/n \geq 2\).}
    \State \textbf{Sample \((\sigma_\epsilon^2)^{(s)}\)}:
    \[
    (\sigma_\epsilon^2)^{(s)} \sim \text{Inv-Gamma}\left( \frac{n + D}{2}, \frac{1}{2} \left( \sum_{i=1}^n \left( y_i - \mathbf{z}_i^T \boldsymbol{\beta}^{(s)} - \beta_0^{(s)} \right)^2 + \sum_{j=1}^D \frac{(\beta_j^{(s)})^2}{(\tau^2)^{(s-1)}} \right) \right).
    \]
    \State \textbf{Sample \((\tau^2)^{(s)}\)}:
    \[
    (\tau^2)^{(s)} \sim \text{Inv-Gamma}\left( \frac{D+1}{2}, \frac{1}{\xi^{(s-1)}} + \frac{1}{2 (\sigma_\epsilon^2)^{(s)}} \sum_{j=1}^D (\beta_j^{(s)})^2 \right).
    \]
    \State \textbf{Sample \(\xi^{(s)}\)}:
    \[
    \xi^{(s)} \sim \text{Inv-Gamma}\left( 1, 1 + \frac{1}{(\tau^2)^{(s)}} \right).
    \]
\EndFor
\State \Return Samples \(\{\beta_0^{(s)}, \boldsymbol{\beta}^{(s)}, (\sigma_\epsilon^2)^{(s)}, (\tau^2)^{(s)}, \xi^{(s)}\}_{s=B+1}^S\).
\end{algorithmic}
\end{algorithm}

\section{Gibbs Sampling for Bayesian Lasso with Gaussian Errors}
\setcounter{algorithm}{0}
According to \citet{makalic2016high}, the Gibbs Sampling for Lasso regression with Gaussian errors under the Bayesian framework is summarized below.

\label{app:bayesian_lasso_gibbs}
\begin{algorithm}[H]
\caption{Gibbs Sampling for Bayesian Lasso with Gaussian Errors}
\label{alg:bayesian_lasso_gibbs}
\begin{algorithmic}[1]
\State \textbf{Input}: Data \(\mathbf{y} \in \mathbb{R}^n\), predictor matrix \(\mathbf{Z} \in \mathbb{R}^{n \times D}\), number of samples \(S\), burn-in \(B\).
\State \textbf{Output}: Posterior samples \(\{\beta_0^{(s)}, \boldsymbol{\beta}^{(s)}, (\sigma_\epsilon^2)^{(s)}, (\lambda_1^2)^{(s)}, \ldots, (\lambda_D^2)^{(s)}, (\tau^2)^{(s)}, \xi^{(s)}\}_{s=B+1}^S\).
\State \textbf{Initialize}:
\State \quad \(\beta_0^{(0)} \), \(\boldsymbol{\beta}^{(0)}\), \((\sigma_\epsilon^2)^{(0)}\), \((\lambda_j^2)^{(0)}\) for \(j = 1, \ldots, D\), \((\tau^2)^{(0)}\), \(\xi^{(0)}\).
\For{\(s = 1, \ldots, S\)}
    \State \textbf{Sample \(\beta_0^{(s)}\)}:
    \[
    \beta_0^{(s)} \sim \mathcal{N}\left( \frac{1}{n} \sum_{i=1}^n \left( y_i - \mathbf{z}_i^T \boldsymbol{\beta}^{(s-1)} \right), \frac{(\sigma_\epsilon^2)^{(s-1)}}{n} \right).
    \]
    \State \textbf{Sample \(\boldsymbol{\beta}^{(s)}\)}:
    \[
    \boldsymbol{\beta}^{(s)} \sim \mathcal{N}_D\left( \tilde{\mu}, \tilde{\boldsymbol{\Lambda}}_D^{-1} \right),
    \]
    \[
    \tilde{\boldsymbol{\Lambda}}_D = \frac{1}{(\sigma_\epsilon^2)^{(s-1)}} \mathbf{Z}^T \mathbf{Z} + \frac{1}{(\sigma_\epsilon^2)^{(s-1)} (\tau^2)^{(s-1)}} \text{diag}\left( \frac{1}{(\lambda_1^2)^{(s-1)}}, \ldots, \frac{1}{(\lambda_D^2)^{(s-1)}} \right),
    \]
    \[
    \tilde{\mu} = \tilde{\boldsymbol{\Lambda}}_D^{-1} \left( \frac{1}{(\sigma_\epsilon^2)^{(s-1)}} \mathbf{Z}^T \left( \mathbf{y} - \beta_0^{(s)} \mathbf{1}_n \right) \right).
    \]
    \Comment{Use \citet{rue2001fast} if \(D/n < 2\), \citet{bhattacharya2016fast} if \(D/n \geq 2\).}
    \State \textbf{Sample \((\sigma_\epsilon^2)^{(s)}\)}:
    \[
    (\sigma_\epsilon^2)^{(s)} \sim \text{Inv-Gamma}\left( \frac{n + D}{2}, \frac{1}{2} \left( \sum_{i=1}^n \left( y_i - \mathbf{z}_i^T \boldsymbol{\beta}^{(s)} - \beta_0^{(s)} \right)^2 + \sum_{j=1}^D \frac{(\beta_j^{(s)})^2}{(\tau^2)^{(s-1)} (\lambda_j^2)^{(s-1)}} \right) \right).
    \]
    \State \textbf{Sample \((\lambda_j^2)^{(s)}\) for \(j = 1, \ldots, D\)}:
    \[
    \left(\frac{1}{\lambda_j^2}\right)^{(s)} \sim \text{IGauss}\left(\left({\frac{2 (\tau^2)^{(s-1)} (\sigma_\epsilon^2)^{(s)}}{(\beta_j^{(s)})^2}}\right)^{1/2}, 2\right), \quad \left(\lambda_j^2\right)^{(s)} = \left[\left( \frac{1}{\lambda_j^2} \right)^{(s)}\right]^{-1}.
    \]
    \State \textbf{Sample \((\tau^2)^{(s)}\)}:
    \[
    (\tau^2)^{(s)} \sim \text{Inv-Gamma}\left( \frac{D+1}{2}, \frac{1}{\xi^{(s-1)}} + \frac{1}{2 (\sigma_\epsilon^2)^{(s)}} \sum_{j=1}^D \frac{(\beta_j^{(s)})^2}{(\lambda_j^2)^{(s)}} \right).
    \]
    \State \textbf{Sample \(\xi^{(s)}\)}:
    \[
    \xi^{(s)} \sim \text{Inv-Gamma}\left( 1, 1 + \frac{1}{(\tau^2)^{(s)}} \right).
    \]
\EndFor
\State \Return Samples \(\{\beta_0^{(s)}, \boldsymbol{\beta}^{(s)}, (\sigma_\epsilon^2)^{(s)}, (\lambda_1^2)^{(s)}, \ldots, (\lambda_D^2)^{(s)}, (\tau^2)^{(s)}, \xi^{(s)}\}_{s=B+1}^S\).
\end{algorithmic}
\end{algorithm}
\bibliographystyle{elsarticle-harv} 
\bibliography{rfBLT_ref}
\end{document}